\documentclass[a4paper,11pt]{article}

\pdfoutput=1

\usepackage[utf8]{inputenc}
\usepackage{amsmath}
\usepackage{amssymb}
\usepackage{amsfonts}
\usepackage{mathrsfs}
\usepackage{graphicx}
\usepackage{booktabs,adjustbox}
\usepackage{caption}
\usepackage{slashed}
\usepackage{verbatim}
\usepackage{float}
\usepackage{setspace}
\usepackage{subfig}
\usepackage{jheppub}

\usepackage{color}
\usepackage{ulem}

\usepackage{bookmark}
\usepackage{wasysym}

\makeatletter
\newcommand{\thickhline}{%
	\noalign {\ifnum 0=`}\fi \hrule height 1pt
	\futurelet \reserved@a \@xhline
}
\makeatother

\usepackage[utf8]{inputenc}
\usepackage{float}
\allowdisplaybreaks[0]

\title{\boldmath  Dark matter freeze-in via a light   fermion mediator: forbidden decay and scattering}

\author[a]{Shao-Ping Li}

\affiliation[a]{Institute of High Energy Physics, Chinese Academy of Sciences, Beijing 100049, China}
 \emailAdd{spli@ihep.ac.cn}

\abstract{The connection between a hidden nonthermal sector and a  thermal plasma can be established by a light thermal fermion mediator.  
	When the  fermion mediator   is much lighter than  the hidden species, kinematically forbidden decay of the mediator can  be opened at finite temperatures to produce the hidden species. Unlike   bosons having quartic couplings, renormalizable forbidden fermion decay generically shares the    same order of couplings with the scattering. We present a dedicated investigation into  the   freeze-in dark matter production via a  thermal  fermion mediator. We demonstrate that   the plasma-induced  decay rate   differs from that calculated via the   tree-level amplitude, but the former can be   obtained from the latter  via constant rescaling.  Furthermore, we find that the relative effect of the forbidden decay and the scattering on the dark matter relic density can be simply estimated via the thermal coupling between the plasma and the mediator. Applying to  different thermal interactions,   we show that  the  forbidden decay contribution  can  reach the level of $4\%\textendash 45\%$ for a thermal coupling at $0.1\textendash 1$.
 
}

\begin{document}
	\maketitle
	\flushbottom
	
%-----------------------------------------------------------------------------------------------------------------------
\section{Introduction}
\label{sec:intro}
 %-----------------------------------------------------------------------------------------------------------------------

A hidden nonthermal species  can be created in the early universe from the thermal plasma   via a light mediator~\cite{Chu:2011be}.  If the hidden sector consists of feebly interacting   dark matter (DM),  the   direct DM detection  could be   challenging. However,  a light mediator  connecting the  DM  with the standard model (SM) can provide  a striking avenue to test the  feeble DM scenarios if the connection between the mediator and the SM is relatively strong and/or the mediator is relatively light~\cite{Hambye:2018dpi}.

 DM production via mediators have received great interests over the past years.  For instance, the millicharged DM production from a vector mediator~\cite{Davidson:2000hf,Chang:2018rso,Dvorkin:2019zdi,Dvorkin:2020xga} and the sterile neutrino DM production via a  scalar mediator~\cite{Kusenko:2006rh,Petraki:2007gq,Merle:2013wta,Adulpravitchai:2014xna,Drewes:2015eoa}. 
There are also   interesting DM scenarios via  a fermion mediator.  A typical example is that the sterile  neutrino itself can be the  mediator to connect a stable dark sector with the SM particles~\cite{Falkowski:2009yz,Gonzalez-Macias:2016vxy,Batell:2017rol,Bandyopadhyay:2018qcv,Becker:2018rve, Folgado:2018qlv,Bandyopadhyay:2020qpn,Biswas:2021kio,Coy:2021sse,Barman:2022scg,Li:2022bpp}.

The phenomenology of DM production  via  mediators is fruitful. The annihilation from DM to the mediator  could generate  secondary fluxes consisting of SM particles via subsequent mediator decay~\cite{Arguelles:2019ouk,Miranda:2022kzs}.  If the  mediator is sufficiently light, it can   contribute to the energy density of the early universe, thereby leaving   imprints in the epochs detectable by the  big bang nucleosynthesis and the cosmic microwave background~\cite{Planck:2018vyg}.  
%In particular, the light mediator may produce an  $N_{{\rm eff}}$ excess which can be probed in future experiments~\cite{SPT-3G:2014dbx,CMB-S4:2016ple,SimonsObservatory:2019qwx}.
Furthermore,
if the  mediator is   long-lived, it can generate displaced vertices and could be detected at the LHC~\cite{Helo:2013esa,Curtin:2018mvb,Belanger:2018sti,Alimena:2019zri}. 
%A neutral fermion coupling to  charged-lepton  and charged-scalar singlets can be   produced by the electron-positron pair, and if the mediator is lighter than the exotic scalar, the opposite-sign dilepton can be produced with missing energy from the charged scalar decay~\cite{Bai:2014osa}.

Generally, if the mediator particle is heavier than the dark sector in vacuum, the mediator decay to DM   plays the dominant role in generating the DM relic density, unless there is   sufficient mass mixing between the mediator and the DM~\cite{Dodelson:1993je}. 
%while the scattering effect is usually subdominant or negligible due to the suppression of higher-order weak couplings and additional phase-space factors. 
If the mediator is much lighter than the dark sector, the decay  channel is kinematically forbidden in vacuum and the scattering/annihilation would be naturally considered as the dominant production channel.  In this latter case, however,
if the mediator has a strong connection with the thermal particles, the mediator  will acquire non-negligible corrections from the    plasma and such   thermal corrections can open up DM production channels which are kinematically forbidden in vacuum. 

Forbidden channels in generating the observed DM relic density were considered  in the thermal freeze-out paradigm~\cite{Griest:1990kh,DAgnolo:2015ujb}, where the   relic density is determined by  the DM annihilation  channel forbidden in vacuum. 
%If the mediator is heavier than the dark sector, such a temperature-dependent mass  effect  is expected to  give a subdominant   correction to the  zero-temperature decay rate.
%However, when the mediator is much lighter than the dark sector, the  temperature-dependent mass  enables a purely plasma-induced decay  which is  kinematically forbidden  in vacuum.  In this case, it is found that the rates from   the scattering/annihilation of  thermal particles and the forbidden mediator decay are at the same order of coupling constants~\cite{Li:2022rde}. Therefore, a consistent treatment from both scattering and the forbidden decay channels is needed to obtain a precise DM relic density.
% Nevertheless, a strong scattering effect is generically expected compared to the forbidden decay, unless the coupling between the mediator and the SM  is sufficiently large.We will illustrate in this work that this conclusion also holds for a fermion mediator.
In the freeze-in paradigm~\cite{McDonald:2001vt,Kusenko:2006rh,Petraki:2007gq,Hall:2009bx,Bernal:2017kxu},   forbidden     decay was considered in Refs.~\cite{Strumia:2010aa,Rychkov:2007uq,Dvorkin:2019zdi,Chang:2019xva} for a vector mediator and in Refs.~\cite{Drewes:2015eoa,Darme:2019wpd,Konar:2021oye,Li:2022rde} for a scalar mediator.  For a fermion mediator,  on the other hand,   the mediator heavier than the DM is usually considered so that the forbidden     decay  contributes only as a subdominant or negligible correction to the vacuum decay (see e.g.~\cite{Gonzalez-Macias:2016vxy,Becker:2018rve,Barman:2022scg}), while few attention  is drawn to  the light mass regime where    forbidden decay and scattering could  coexist to generate the   relic density. Filling in this gap underlies the purpose of this work.

%When the fermion mediator is much lighter than the DM, the forbidden decay is no longer expected to be a subdominant correction to the vacuum decay. 
A thermal fermion mediator differs by several aspects from a boson mediator. Since the boson  can have a renormalizable self-interaction, such as the gluons  and an SM scalar singlet, the forbidden decay rate can carry  lower-order couplings with respect to the scattering so that the former becomes the dominant production channel~\cite{Rychkov:2007uq,Darme:2019wpd,Konar:2021oye}, unless the quartic interactions are suppressed with respect to Yukawa or trilinear-boson interaction~\cite{Li:2022rde}.   For fermion mediators,  renormalizable interactions imply that there is no tree-level quartic fermion interaction, and the forbidden fermion  decay rate  would generically  carry the same order of   couplings with respect to the scattering, as will shown in this work. Given that  the rates  from the forbidden   decay and the scattering have the same order of couplings,  it becomes less clear  to see  the relative effect of the  forbidden   decay and the associated scattering on the DM relic density and hence  worth examining in detail the interplay between the two channels.

 On the other hand,  the modified dispersion relation of a   scalar  at finite temperatures  retains the vacuum form
%~\cite{Comelli:1996vm}   function~ 
in the Hard-Thermal-Loop (HTL) approximation~\cite{Braaten:1989mz,Frenkel:1989br,Braaten:1991gm,Carrington:1997sq,Bellac2000}, which allows the calculation of forbidden scalar decay to follow  a tree-level amplitude~\cite{Drewes:2015eoa,Li:2022rde}.  For fermion mediators, however, the modified dispersion relation is more involved due to the helicity structure~\cite{Weldon:1982bn,Braaten:1990wp,Peshier:1999dt}. It then becomes nontrivial to see if and how the forbidden decay can be simply obtained via a tree-level amplitude, where the thermal fermion mass is put in by hand. Such an issue was considered   in leptogenesis~\cite{Kiessig:2010pr} and in this paper, we bring it for the first time to the freeze-in DM production and provide a comprehensive analysis on the difference between the tree-level  and one-loop results.

This work is concerned with    a dedicated  analysis of freeze-in  DM production via  a light thermal   fermion mediator which cannot decay to   DM at zero temperature.
We will  concentrate on the computation of the DM relic density from the forbidden fermion decay and the scattering.
We will calculate the forbidden decay rates from a thermal one-loop amplitude and a vacuum tree-level amplitude, respectively, and find that  the former can be simply obtained from the latter with some constant rescaling.
The comparison between the  forbidden decay and the scattering shows a rather simple dependence on the thermal coupling constant, which enables us to include the plasma-induced decay in the scattering channel  efficiently.
 This work complements the studies of nonthermal DM production through a light fermion mediator and provides a simple and comprehensive method to treat the forbidden   decay for a wide range of fermion mediator scenarios. In particular, the results shown here can help us to gain a clear insight into the importance of forbidden fermion decay.

The remainder of this paper is outlined as follows.  In Sec.~\ref{sec:simplemodel}, we present a simplified but general scenario to illustrate the freeze-in DM production via a light   thermal fermion mediator. Within the simplified scenario, we calculate the forbidden decay rate in Sec.~\ref{sec:forbiddendecay} and make a comparison with the rate derived from the vacuum tree-level amplitude.  In Sec.~\ref{sec:scattering}, we first   point out some subtleties concerning the double-counting issue and the  $s$-channel resonant enhancement,  and then  evaluate the scattering rate without thermal corrections.   In Sec.~\ref{sec:relicdensity}, we determine the DM relic density   from the forbidden decay and scattering channels respectively. 
%The relative contribution can be simply estimated via the thermal coupling between  the mediator and the thermal plasma. 
 We then apply the relation between the two channels to some specific thermal interactions in Sec.~\ref{sec:UVmodels}. 
Conclusions are   made in Sec.~\ref{sec:conclusion} and some technical details are relegated to the appendix.

%-----------------------------------------------------------------------------------------------------------------------
\section{The  Yukawa portals}\label{sec:simplemodel}
%-----------------------------------------------------------------------------------------------------------------------
We first consider a simplified scenario in which the nonthermal dark sector consists of a  Dirac fermion $\chi$ and a scalar $\phi$. The connection between the  dark sector  and  a   Dirac fermion  mediator $\psi$ is realized by the following Yukawa interaction:
\begin{align}\label{eq:Lag}
\mathcal{L_{{\rm DM}}}=y_{\chi}\bar{\psi}_{R}\chi_{L}\phi+\rm h.c.
\end{align}
To ensure a thermal history of $\psi$, we  consider a typical Yukawa interaction between the mediator and the thermal plasma, i.e.,
\begin{align}\label{eq:Yukawathermal}
\mathcal{L}_{\psi}	=y_{\psi}\bar{\psi}_{R}\eta_{L}\varphi+\rm h.c.\, ,
\end{align} 
where both the fermion $\eta$ and the scalar $\varphi$ live in the thermal plasma.  For clarity, we assume that the fermion mediator is   right-handed in~\eqref{eq:Lag}, but it should be mentioned that a left-handed fermion mediator is also possible. 
In Sec.~\ref{sec:UVmodels}, we shall discuss some   specific models  for both right- and left-handed fermion mediators.

 Note that the fermion mediator can also have gauge interactions, e.g., 
 \begin{align}
V_{\mu}\bar{\psi}_{R}\gamma^{\mu}\psi_{R}\, ,
 \end{align}
with $V_\mu$ a $U(1)$ gauge boson. Nevertheless, when the mediator is thermalized via the gauge interaction, gauge invariance requires that either $\chi$ or $\phi$ should be also charged under the gauge $U(1)$ symmetry. In this case,  either $\chi$ or $\phi$ will   reach thermal equilibrium in the early universe, which can lead to  significant difference from the situation where both $\chi$ and $\phi$ are far from equilibrium. For instance, when $\phi$ is in thermal equilibrium, the decay $\phi\to \chi+\psi$ and the scattering $\phi+\psi\to \chi+V_\mu$ can dominate the production of $\chi$, both of which are suppressed instead when $\phi$ is far from equilibrium. Besides, the Landau-Pomeranchuk-Migdal effect induced by soft  vector boson exchange
would also be of leading-order contribution~\cite{Arnold:2002zm} and should be taken into account consistently.
Throughout this work, we will consider for simplicity a dark sector consisting of nonthermal $\chi$ and $\phi$, leaving a thermal $\chi$ or $\phi$ for future studies.

%------------------------------------------------------------------------------------------------------
\begin{figure}[t]
	\centering
	\includegraphics[width=0.3\textwidth]{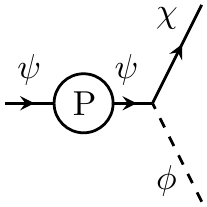}
	\caption{ Forbidden fermion decay due to the thermal contact with the plasma (P).
		\label{fig:SVSD}
	}
\end{figure}
%------------------------------------------------------------------------------------------------------

We will consider the situation where all the relevant thermal particles, i.e., $\psi$, $\eta$, and  $\varphi$  have vacuum masses much lighter than the dark sector, which is  readily applicable to super-heavy DM~\cite{Chung:1998zb,Chung:2001cb}. In this light mass regime, the freeze-in temperature  of the DM is determined by the highest scale in the dark sector. Besides,   without a mild mass difference between the initial and final states, as implemented in forbidden annihilation channels~\cite{Griest:1990kh,DAgnolo:2015ujb},  the nonrelativistic  annihilation of $\psi$, $\eta$, and $\varphi$ to the dark sector is essentially disallowed. Consequently,  the DM relic density  would basically be independent of the vacuum masses of the thermal particles. In the following discussions, 
we assume $m_\chi<m_\phi$ for clarity. In this mass regime, either $\chi$ can be   the only DM candidate or both $\chi$ and $\phi$ contribute to  the observed DM relic density, though the later case is ruled out if $m_\phi\gg 1$~GeV.

Before going into the detailed calculation, let us take a diagrammatic view of the relation between the forbidden fermion decay and the scattering. In Fig.~\ref{fig:SVSD}, the  fermion $\psi$ receives a thermal mass correction from  the self-energy diagram, where P denotes the plasma. Such a correction opens the kinematic space  so that the decay $\psi\to \chi+\phi$ becomes possible at finite temperatures. Dimensional analysis implies that the squared amplitude scales as $y_\chi^2m_\psi^2$ at   sufficiently high temperatures, where  $m_\psi$ denotes the thermal mass. The interaction given in~\eqref{eq:Yukawathermal} implies that $m_\psi^2\sim y_\psi^2T^2$. Therefore, the forbidden decay rate at high temperatures scales as
\begin{align}
	\gamma_{\rm decay}\sim y_\chi^2 y_\psi^2T^4\, .
\end{align}
The thermal self-energy amplitude in general has an imaginary part, which corresponds to on-shell thermal particles running in the loop. In this case,  Fig.~\ref{fig:SVSD} also presents a scattering channel $\eta+\varphi\to \psi\to \chi+\phi$. It is easy to see that the squared amplitude of the scattering also depends on    $y_\chi$ and $y_\psi$ quadratically, and  the scattering rate at high temperatures would  scale as
\begin{align}
	\gamma_{\rm scat}\sim y_\chi^2 y_\psi^2 T^4\sim \gamma_{\rm decay}\, .
\end{align}
It should be mentioned that the self-energy correction for relativistic fermions generically predicts a thermal mass with a form $\sim y T$, where $y$ is the dimensionless coupling between the fermion mediator and the plasma\footnote{We are concerned with IR-dependent freeze-in so that the production of DM comes from renormalizable interactions. The conclusions drawn in this paper are hence responsible  for renormalizable interactions. For non-renormalizable interactions, the freeze-in production of DM is not IR but UV dependent~\cite{Elahi:2014fsa}. }. Such a  fermion mediator in renormalizable interactions differs from a vector/scalar boson mediator which has a strong quartic self-coupling $\lambda$.  The leading-order thermal mass for such bosons scales as $\sim \sqrt{\lambda} T$ and the resulting forbidden boson decay has a rate   $	\gamma_{\rm decay}\propto \lambda T^4$ while the associated scattering  rate gives $\gamma_{\rm scat}\propto \lambda^2 T^4$. In such situations, the forbidden decay can dominant the DM production, as considered in Refs.~\cite{Rychkov:2007uq,Darme:2019wpd,Konar:2021oye}.

Therefore, unlike the forbidden boson decay, the scattering is   present at the same order of couplings  whenever forbidden fermion decay is opened in renormalizable interactions, and vice versa. They  coexist to produce the DM at finite temperatures. We will show in the subsequent content that there is a close relation between the two channels, which allows us to see the relative effect on the DM production in a simple way.

%-----------------------------------------------------------------------------------------------------------------------
\section{Forbidden decay}\label{sec:forbiddendecay}
%-----------------------------------------------------------------------------------------------------------------------
%-----------------------------------------------------------------------------------------------------------------------
\subsection{Boltzmann equation}
%-----------------------------------------------------------------------------------------------------------------------
The decay process $\psi \to \chi+\phi$ is kinematically forbidden in vacuum but opened at finite temperatures. The forbidden decay rate that determines the density evolution   in the dark sector can be calculated in the finite-temperature field theory~\cite{Bellac2000}. Concerning the production of $\chi$, the   Boltzmann equation  can be written as
\begin{align}\label{eq:Boltzmann}
	\frac{d n_\chi}{d t}+3Hn_{\chi}=\int\frac{d^{3}p_{\chi}}{(2\pi)^{3}}(f_{\chi}^{{\rm eq}}-f_{\chi})\Gamma_{\chi}\, ,
%	\nonumber \\[0.2cm]
%&\approx-g_{\chi}\int\frac{d^{3}p_{\chi}}{(2\pi)^{3}2E_{\chi}}f_{\chi}^{{\rm eq}}\text{Tr}[(\slashed{P}+m_{\chi})\text{Im}\Sigma_{R}^{\chi}(P)],
\end{align}
where $f_\chi^{\rm eq}(E_\chi)=(e^{E_\chi/T}+1)^{-1}$ is the Fermi-Dirac distribution function of $\chi$  and $H\approx 1.66 \sqrt{g_\rho}T^2/M_{\rm Pl}$ is the Hubble parameter  with the effective number of relativistic degrees of freedom $g_\rho$ for energy density and the Planck mass $M_{\rm Pl}\approx 1.22\times 10^{19}$~GeV. 

The   production rate $\Gamma_{\chi}$ at finite temperatures  is related to the  one-loop retarded  self-energy of $\chi$ via~\cite{Weldon:1983jn}
 \begin{align}\label{eq:Gammachi}
	\Gamma_{\chi}(P)=-g_{\chi}\frac{\text{Tr}[(\slashed{P}+m_{\chi})\text{Im}\Sigma_{R}^\chi(P)]}{2E_p}\, ,
\end{align}
with $P_{\mu}=(E_p,\vec{p})$ the 4-momentum of $\chi$ and $\text{Im}\Sigma_{R}^\chi$ the imaginary part of the  one-loop retarded amplitude.
It should be mentioned that the factor of 2 in the denominator of Eq.~\eqref{eq:Gammachi} results from the spin sum and average over the Dirac spinor $\chi$. Therefore, the collision rate in the Boltzmann Eq.~\eqref{eq:Boltzmann} should be further multiplied by the spin degrees of freedom $g_\chi=2$~\cite{Li:2022dkc} so as to obtain a collision term without spin average.  For a nonthermal DM in the freeze-in paradigm, we expect $f_{\chi}\ll f_{\chi}^{\rm eq}$ so that  $f_{\chi}$ can be neglected in the  determination of   the DM relic density. In the end, the relic density should   be multiplied by a factor of 2 to take into account  the antiparticle ($\bar{\chi}$) contribution.

%------------------------------------------------------------------------------------------------------
\begin{figure}[t]
	\centering
	\includegraphics[width=0.96\textwidth]{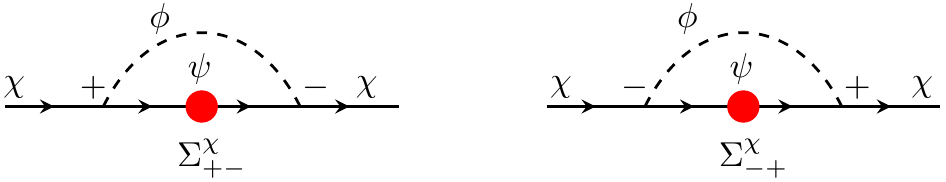}
	\caption{The one-loop self-energy diagrams of $\chi$ that contribute to the imaginary part of the retarded amplitude $\text{Im}\Sigma_{R}^\chi$ in the forbidden decay. Here $\pm$ in the vertices denote the thermal indices in the doubled space of real-time formalism and   the red blob denotes the resummed $\psi$ propagator at finite temperatures.  
		\label{fg:SigmaRchi}
	}
\end{figure}
%------------------------------------------------------------------------------------------------------

In the real-time formalism, the imaginary part of the retarded amplitude $\Sigma_{R}^\chi$  can be computed  from the one-loop self-energy diagrams shown in Fig.~\ref{fg:SigmaRchi},  with
%onvention~\cite{Carrington:1997sq, Thoma:2000dc}, 
%$\text{Im}\Sigma_{R}^{\chi}(K)$ can be computed via 
 \begin{align}
\text{Im}\Sigma_{R}^{\chi}(P)=\frac{i}{2}\left[\Sigma_{+-}^{\chi}(P)-\Sigma_{-+}^{\chi}(P)\right].
\end{align}
 Using the  expressions of $\Sigma_{+-}^{\chi},\Sigma_{-+}^{\chi}$ from Appendix~\ref{appendix:A1}, we obtain
 \begin{align}\label{eq:SigmachiR}
\text{Im}\Sigma_{R}^{\chi}(P)=\frac{y_{\chi}^{2}}{2(2\pi)^{2}}\int d^{4}K\text{sign}(k_{0}-p_{0})f_{\psi}(k_{0})\delta[(K-P)^{2}-m_{\phi}^{2}]\rho_{\psi}(K)\, ,
\end{align}
where $\text{sign}(k_0-p_0)$ denotes the sign function and $f_\psi(k_0)=(e^{k_0/T}+1)^{-1}$.  In the above equation, we have neglected the scalar distribution function $f_\phi$ since $\phi$ is sparse during the freeze-in production.   $\rho_\psi(K)$ is the  spectral density that encapsulates  the thermal corrections to $\psi$, as we shall derive below.

%-----------------------------------------------------------------------------------------------------------------------
\subsection{Spectral density of the fermion mediator}\label{sec:spectraldensity}
%-----------------------------------------------------------------------------------------------------------------------

The spectral density is defined via the resummed $\psi$ propagators,
\begin{align}\label{eq:S+-}
S_{+-}	&=-f_{\psi}(\tilde{G}_{R}-\tilde{G}_{A})\equiv-2\pi if_{\psi}(k_{0})\rho_{\psi}(K)\, ,
\\[0.2cm]
S_{-+}	&=[1-f_{\psi}(k_{0})](\tilde{G}_{R}-\tilde{G}_{A})\equiv2\pi i[1-f_{\psi}(k_{0})]\rho_{\psi}(K)\, , \label{eq:S-+}
\end{align}
where $\tilde{G}_R/\tilde{G}_A$ are the resummed retarded/advanced propagators.
Since the spectral density defined above encapsulates the thermal corrections in the form  of $\tilde{G}_R-\tilde{G}_A$, we should first be aware of    how the  thermal corrections appear in the resummed retarded and advanced propagators.

 In general, the retarded amplitude for fermion self-energy can be parameterized as\footnote{The minus sign is defined for convenience, which results in $1+a$ in the denominator of propagators.}~\cite{Weldon:1982bn}
\begin{align}\label{eq:SigmaRpsi}
	-\Sigma_{R}^{\psi}(K)\equiv(a_{L}P_{L}+a_{R}P_{R})\slashed{K}+(b_{L}P_{L}+b_{R}P_{R})\slashed{U}\, ,
\end{align}
where $P_{L,R}$ are the chirality projection  operators and $U_{\mu}$ is the four-velocity of the plasma with $U_{\mu}U^{\mu}=1$. In the rest frame, $U_{\mu}=(1,0,0,0)$. 
Since the parity of the fermion mediator from the interactions given in Sec.~\ref{sec:simplemodel} is explicitly broken, and at sufficiently high temperatures $\psi$ is effectively massless\footnote{If $\psi$ acquires its vacuum mass via the Higgs or Higgs-like mechanism, then $\psi$ is exactly massless above the cross-over or phase-transition  temperature.}, we are essentially working in  
a chirality-symmetric and  parity-broken theory,  where $a_{L},b_{L}$ are nonzero while $a_{R},b_{R}=0$.    
The coefficients $a_{L},b_{L}$ can be calculated by left-multiplying $\Sigma_{R}^\psi(K)$ with $\slashed K$  and $\slashed U$,  and then evaluating the trace. The general expressions read:
\begin{align}\label{eq:aL}
	a_{L}	&=\frac{1}{2k^{2}}\left(\text{Tr}[\slashed{K}\Sigma_{R}^{\psi}(K)]-k_{0}\text{Tr}[\slashed{U}\Sigma_{R}^{\psi}(K)]\right),
	\\[0.2cm]
	b_{L}	&=-\frac{1}{2k^{2}}\left(k_{0}\text{Tr}[\slashed{K}\Sigma_{R}^{\psi}(K)]-K^{2}\text{Tr}[\slashed{U}\Sigma_{R}^{\psi}(K)]\right),\label{eq:bL}
\end{align}
with $K^{2}=k_{0}^{2}-k^{2}$. 

Given Eq.~\eqref{eq:SigmaRpsi},  the resummed retarded propagator in the  chirality-symmetric and  parity-broken regime can be written as 
\begin{align}
	\tilde{G}_{R}	=P_{R}\frac{(1+a_{L})\slashed{K}+b_{L}\slashed{U}}{[(1+a_{L})k_0+b_{L}]^{2}-[(1+a_{L})k]^{2}+i\text{sign}(k_0)\epsilon}P_{L}\, ,
\end{align}
and the advanced propagator can be similarly obtained by using  $\Sigma_A^\psi=\Sigma_R^{\psi *}$.  The difference $	\tilde{G}_{R}-\tilde{G}_{A}$  can be  conveniently written  in terms of the helicity eigenstates~\cite{Braaten:1990wp,Peshier:1999dt},
\begin{align}\label{eq:GR-GA}
	\tilde{G}_{R}-\tilde{G}_{A}	=\sum_{\pm}\frac{-2i(\text{Im}\Delta_{+}\mp\text{sign}(k_0)\epsilon)}{[\text{Re}\Delta_{\pm}]^{2}+[\text{Im}\Delta_{\pm}+\epsilon]^{2}}\hat{P}_\pm\, ,
\end{align}
where $\Delta_{\pm}(K)\equiv(1+a_{L})k_{0}+b_{L}\pm(1+a_{L})k$, and the helicity operators are defined by
\begin{align}
	\hat{P}_\pm\equiv P_{R}\frac{\gamma^{0}\pm \vec{e}_{k}\cdot\vec{\gamma}}{2}P_{L}\, ,
\end{align}
with $\vec{e}_k\equiv\vec{k}/k$. 

The spectral density $\rho_\psi$ can be decomposed into the on-shell and off-shell parts,
\begin{align}
	\rho_{\psi}(K)\equiv \rho_{\psi,{\rm on}}(K)+\rho_{\psi,{\rm off}}(K)\, .
\end{align}
The  kinematically forbidden decay stems from the on-shell part $\rho_{\psi,{\rm on}}(K)$, as will be derived in this section, while the off-shell part $\rho_{\psi,{\rm off}}(K)$ arises from nonzero $\text{Im}\Delta_{\pm}$ and corresponds to the  scattering channels.  Note that  the on-shell propagation of the fermion mediator could also result from the scattering channel. To avoid potential double counting, $\rho_{\psi,{\rm on}}(K)$ defined above corresponds to $\text{Im}\Delta_{\pm}=0$. Then,
from Eq.~\eqref{eq:GR-GA}  the on-shell part is given by
 \begin{align}\label{eq:rhopsi-on}
\rho_{\psi,{\rm on}}(K)	=
\sum_{\pm}\pm\text{sign}(k_{0})\Big|\frac{\partial\text{Re}\Delta_{\pm}}{\partial k_{0}}\Big|^{-1}\Big[\delta(k_{0}-\omega_{1}^{\pm})+\delta(k_{0}-\omega_{2}^{\pm})\Big]\hat{P}_\pm\, .
\end{align}
In general, there are  two solutions $\omega_{1,2}$ to  $\text{Re}\Delta_{i}=0$
 for each helicity operator $\hat{P}_i$.
In the free limit, $a_L=b_L=0$ and $\Delta_{\pm}=k_{0}\pm k$. It can be verified  that $S_{+-},S_{-+}$  given in Eqs.~\eqref{eq:S+-} and~\eqref{eq:S-+} reduce to the known forms~\cite{Bellac2000}:
\begin{align}\label{eq:freeS+-}
S_{+-}(K)	&=2\pi i\text{sign}(k_{0})f_{\psi}(k_{0})\delta(K^{2})\slashed{K}\, ,
%=2\pi i[-\theta(-k_{0})+f_{\psi}(|k_{0}|)]\delta(K^{2}) \slashed{K}\, ,
\\[0.3cm]
S_{-+}(K)	&=-2\pi i\text{sign}(k_{0})[1-f_{\psi}(k_{0})]\delta(K^{2})\slashed{K}\,.\label{eq:freeS-+}
%=2\pi i[-\theta(k_{0})+f_{\psi}(|k_{0}|)]\delta(K^{2})\slashed{K}\,.
 \end{align}
To proceed with   Eq.~\eqref{eq:SigmachiR}, the remaining task   is to evaluate the real part of the  resummed   amplitude $\Sigma_R^\psi$, which depends on the thermal interaction specified in Sec.~\ref{sec:simplemodel}.

The  one-loop retarded self-energy diagram of $\psi$ from~\eqref{eq:Yukawathermal} is similar to Fig.~\ref{fg:SigmaRchi}, with the resummed fermion propagators replaced by the free ones given in Eqs.~\eqref{eq:freeS+-} and~\eqref{eq:freeS-+}. The inclusion of resummed propagators for   thermal $\eta,\varphi$  in Fig.~\ref{fg:SigmaRchi} is of higher order under the perturbative HTL technique.
Substituting Eqs.~\eqref{eq:trKSigma} and~\eqref{eq:trUSigma} into Eqs.~\eqref{eq:aL} and \eqref{eq:bL}, we obtain the real part of the coefficients $a_L, b_L$ as
 \begin{align}\label{eq:ReaL}
\text{Re}a_{L}	&=\frac{m_{\psi}^{2}(T)}{k^{2}}\left(1+\frac{k_{0}}{2k}\ln\left|\frac{k_{0}-k}{k_{0}+k}\right|\right),
\\[0.2cm]
\text{Re}b_{L}	&=-\frac{m_{\psi}^{2}(T)}{k}\left(\frac{k_{0}}{k}-\frac{1}{2}\left(1-\frac{k_{0}^{2}}{k^{2}}\right)\ln\left|\frac{k_{0}-k}{k_{0}+k}\right|\right) ,\label{eq:RebL}
 \end{align}
where the thermal mass is defined by
\begin{align}\label{eq:thermalmass}
	m_{\psi}^{2}(T)= \frac{y_{\psi}^{2}}{16}T^{2}\equiv \kappa^2 T^2\, ,
\end{align}  
where $\kappa$ is defined as a thermal parameter quantifying the amount of thermal corrections.

The results given in  Eqs.~\eqref{eq:ReaL} and~\eqref{eq:RebL}    are consistent with Ref.~\cite{Weldon:1982bn} except that the logarithmic function  is expressed by the modulus of momentum.  The modulus arises when we   integrate $\cos\theta$ in Eq.~\eqref{eq:trUSigma} without restricting ourselves to the timelike regime $K^2=k_0^2-k^2>0$.  Nevertheless, we will see below that  an  on-shell   fermion with Eqs.~\eqref{eq:ReaL} and~\eqref{eq:RebL} cannot   propagate in the spacelike region.   The  modified    dispersion relation is given by
 \begin{align}\label{eq:fermiondisper}
	\left[(1+\text{Re}a_{L})k_{0}+\text{Re}b_{L}\right]^{2}-\left[(1+\text{Re}a_{L})k\right]^{2}=0\, .
\end{align}
For a weak-coupling theory $y_\psi\lesssim 1$, we expect  $\text{Re}a_{L}<1$. Neglecting the   higher-order terms  $\text{Re}a_{L}^2$ and $\text{Re}b_{L}^2$, we obtain the approximate   dispersion relation:
 \begin{align}\label{eq:fermiondisper2}
k_0^2-k^2\approx -\frac{2k_0 \text{Re}b_{L}}{1+2\text{Re}a_{L}}\, .
\end{align}
Then given  Eqs.~\eqref{eq:ReaL} and \eqref{eq:RebL}, it is straightforward to verify that there is no solution to the above equation for   $k_0^2-k^2<0$.  
Therefore, the absolute symbol in Eqs.~\eqref{eq:ReaL} and \eqref{eq:RebL}  should  be removed. 

The thermal mass defined in Eq.~\eqref{eq:thermalmass} is proportional to the quadratic Casimir invariant of the fermion mediator representation in  gauge interactions, as well as the gauge degeneracy of the loop particles in Yukawa interactions~\cite{Weldon:1982bn}. For the freeze-in DM production considered here,   the fermion mediator   should be a SM singlet so there is no gauge contribution to the  thermal mass. However,  the loop particles could be gauge multiplets. For instance, if the scalar $\varphi$ and the fermion $\eta$  are gauge $SU(2)_L$ doublets,  then an additional  factor of 2 arises in $m_{\psi}^{2}(T)$.  This is readily seen by the fact that there are two gauge components in the loop. On the other hand, if  $\varphi$ and   $\eta$  are  gauge $SU(3)_c$ triplets, a factor of 3  due to the color degrees of freedom arises in $m_{\psi}^{2}(T)$.  

It should also be mentioned that the   results given in Eqs.~\eqref{eq:ReaL}-\eqref{eq:thermalmass} (see also the appendix) are obtained in the HTL approximation which keeps the leading-order  coupling ($y_\psi$) contributions. Under the perturbative region $y_\psi<\sqrt{4\pi}$, there is no definite upper limit of the coupling for the  HTL validity. It was pointed out in Ref.~\cite{Peshier:1998dy} that a coupling at $\sim 1$ can still give  qualitatively correct result under the HTL approximation.   In general, larger couplings lead to a poorer accuracy under the HTL approximation. Therefore, we will  impose  $y_\psi<1$ as a conservative upper limit for the Yukawa interaction when applying the HTL approximation. In particular, we will consider a weak-coupling regime where
\begin{align}\label{eq:weakcoup}
	0.1< y_\psi<1\, ,
\end{align}
in subsequent discussions. 

In the following analyses, we will take $\kappa$ as a free thermal parameter. It it noteworthy that  the  upper bound of $\kappa$ from the condition  in Eq.~\eqref{eq:weakcoup} depends on 
the specific thermalization interaction between the mediator and the plasma, as well as the flavor effects from Yukawa interactions. For instance, if the fermion mediator couples comparably to three SM quark doublets via a leptoquark doublet, the upper bound of $\kappa$ is given by $\kappa<1.1$.  In Sec.~\ref{sec:UVmodels}, we will consider some specific examples and present the corresponding limit of $\kappa$ and its impact on forbidden decay contribution.

%-----------------------------------------------------------------------------------------------------------------------
\subsection{Collision rate }
%-----------------------------------------------------------------------------------------------------------------------
\subsubsection{One-loop retarded amplitude}
Given the expressions of $\text{Re}a_L, \text{Re}b_L$ in Eqs.~\eqref{eq:ReaL} and \eqref{eq:RebL},   the  on-shell spectral density  from Eq.~\eqref{eq:rhopsi-on} can be simplified as
\begin{align}
	\rho_{\psi,{\rm on}}(K)=\sum_{\pm} \pm \frac{k_{0}^{2}-k^{2}}{2m_{\psi}^{2}(T)}\text{sign}(k_0)\left[\delta(k_{0}\mp \omega_{1})+\delta(k_{0}\pm \omega_{2})\right]\hat{P}_{\pm}\,,
\end{align}
where   $\omega_{1,2}$  are the solutions to the modified dispersion relation~\eqref{eq:fermiondisper} and can be analytically expressed in terms of the  Lambert W-function~\cite{Kiessig:2010pr}:
\begin{align}
 \omega_{1}=-k\frac{W_{0}(-e^{-2k^{2}/m_\psi^2-1})-1}{W_{0}(-e^{-2k^{2}/m_\psi^2-1})+1}
\,,\quad \omega_{2}=k \frac{W_{-1}(-e^{-2k^{2}/m_\psi^2-1})-1}{W_{-1}(-e^{-2k^2/m_\psi^2-1})+1}\, ,
\end{align}
with $\omega_{1,2}>k$.

%------------------------------------------------------------------------------------------------------
\begin{figure}[t]
	\centering
	\includegraphics[width=0.49\textwidth]{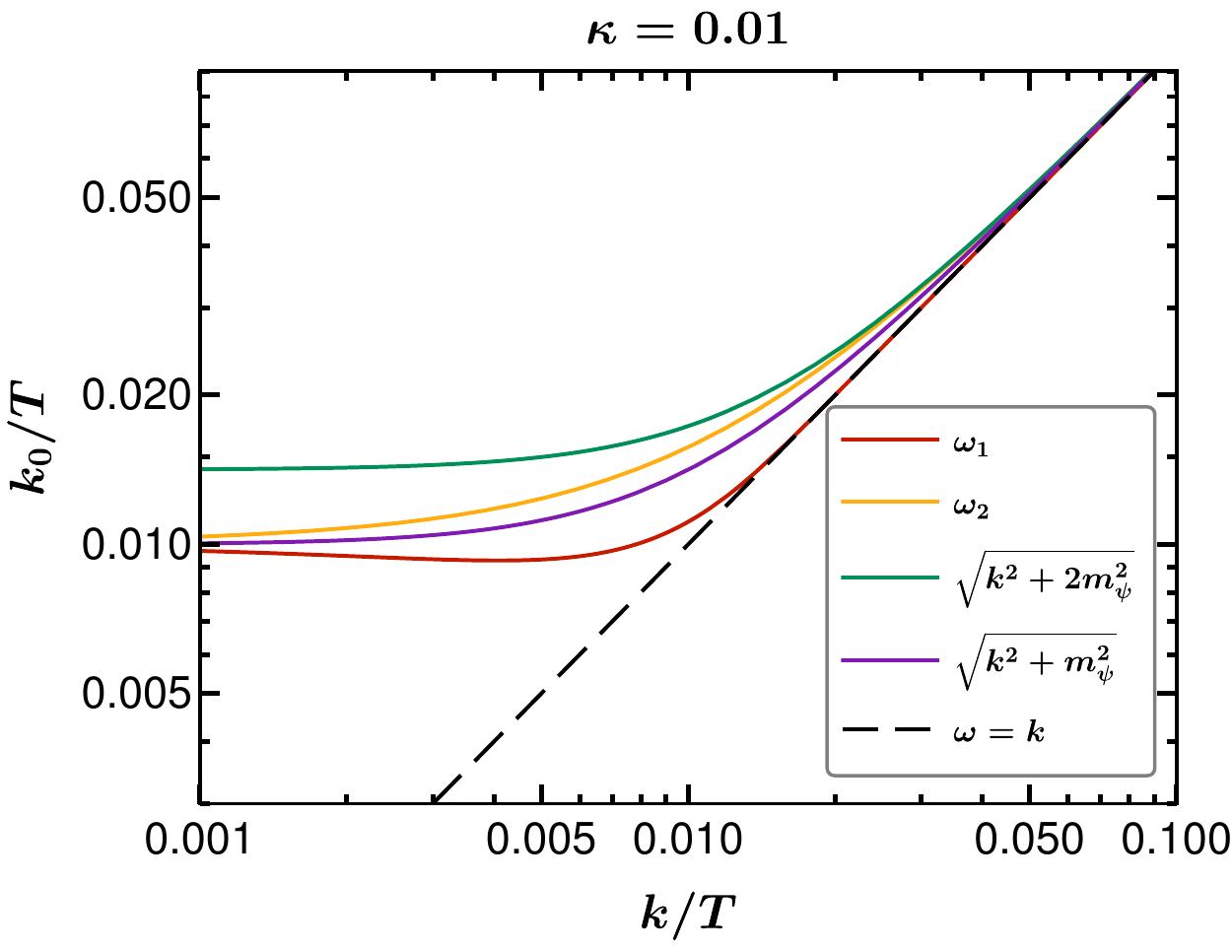}
	\includegraphics[width=0.47\textwidth]{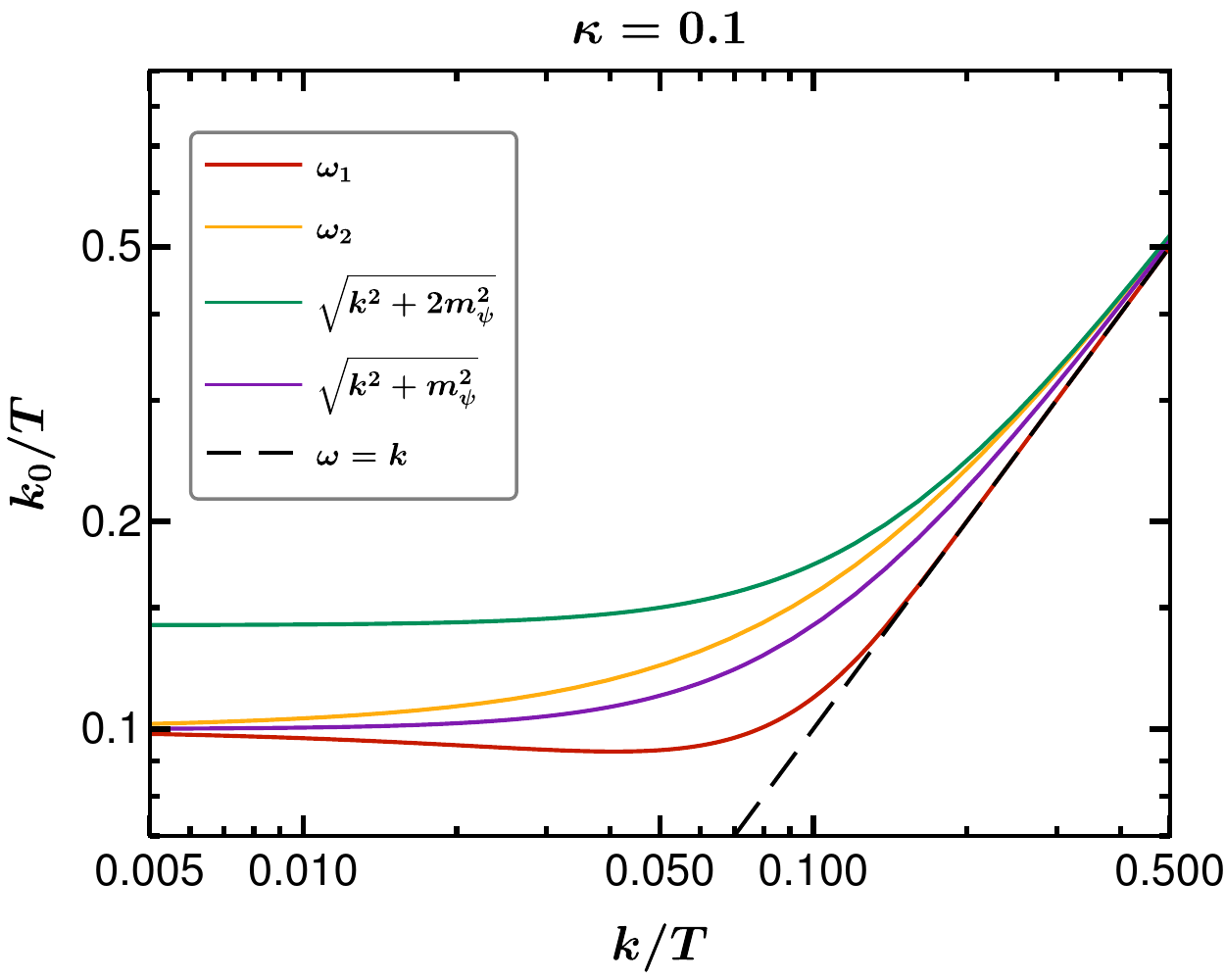}
	\\
	\includegraphics[width=0.49\textwidth]{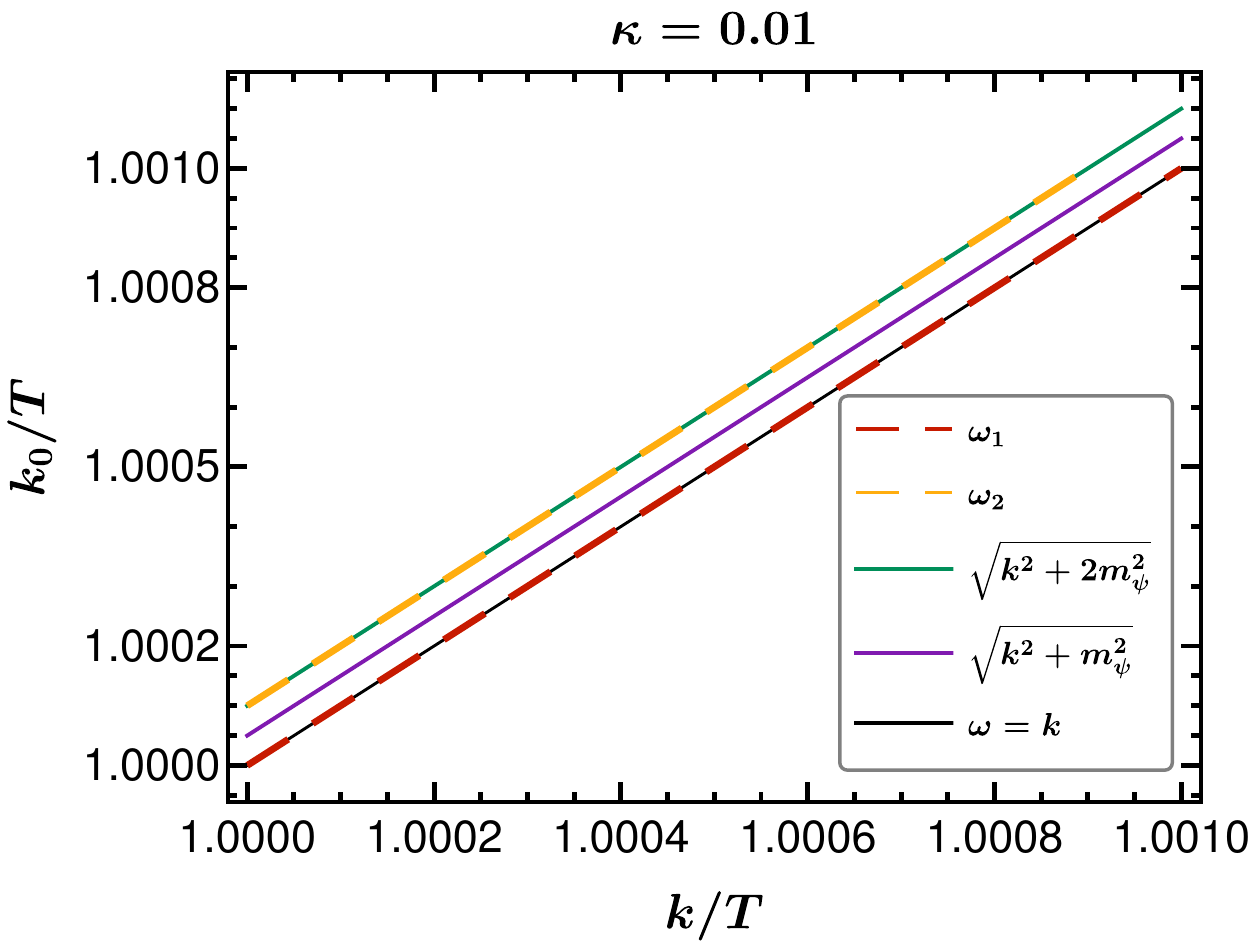}
	\includegraphics[width=0.47\textwidth]{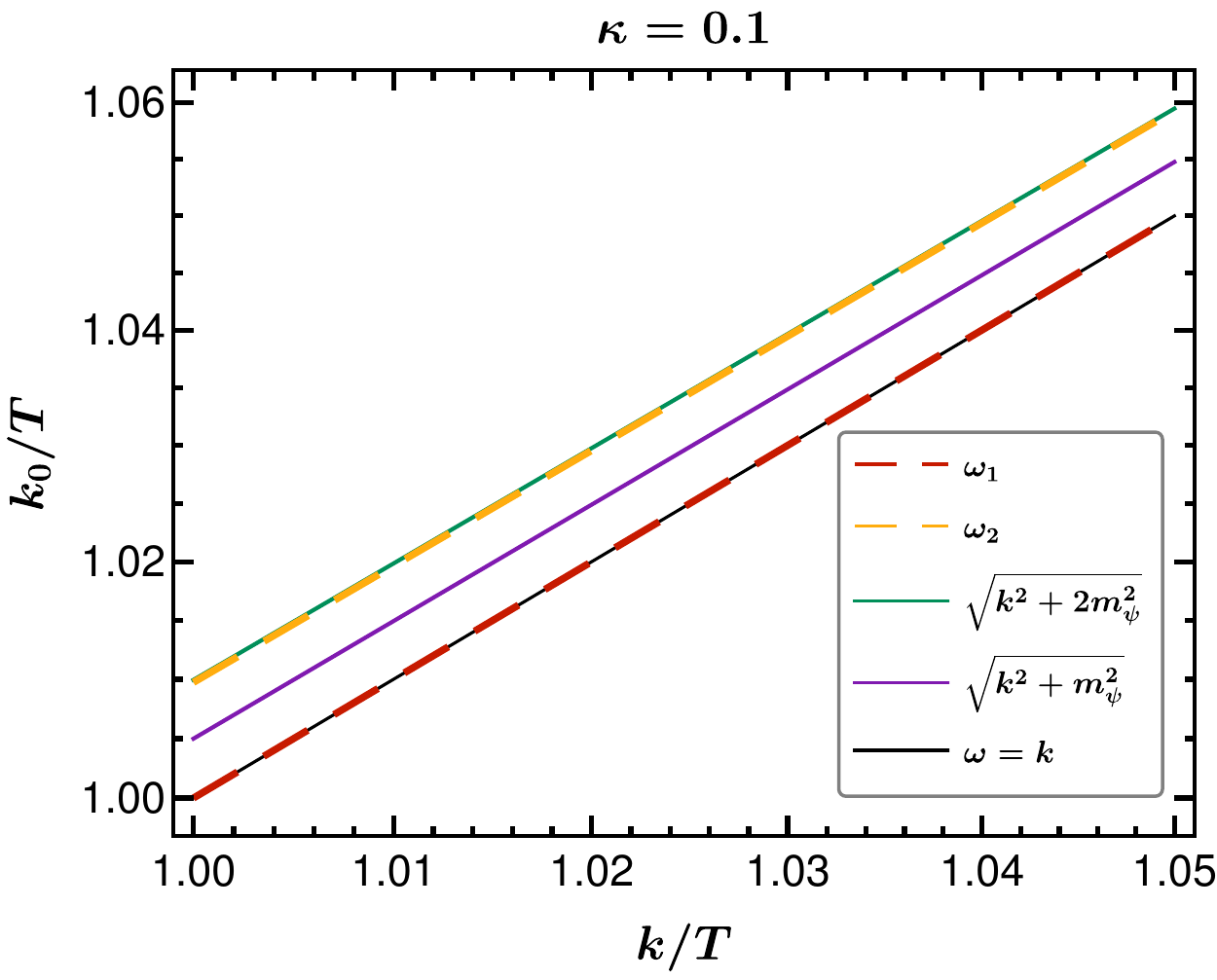}
	\caption{The behavior of dispersion relation~\eqref{eq:fermiondisper} at  $k/T<1$ (top) and $k/T>1$ (bottom), where the thermal parameter is set by $\kappa=0.01, 0.1$, respectively.
		\label{fig:dispersion}
	}
	
\end{figure}
%------------------------------------------------------------------------------------------------------

 Substituting  Eqs.~\eqref{eq:SigmachiR} and \eqref{eq:Gammachi} into the collision term  in   Eq.~\eqref{eq:Boltzmann}, 
we arrive at  the decay rate
\begin{align}\label{eq:Cchidecay}
C_{\chi,{\rm dec}}&=\frac{y_{\chi}^{2}}{32\pi^{3}m_{\psi}^{2}(T)}\int_{m_{\chi}}^{\infty}dp_{0}f_{\chi}^{{\rm eq}}(p_{0})
\nonumber \\[0.2cm]
&\times \int_0^\infty dk\sum_{i=1,2}  \mp \Theta_{i}(\omega_{i}^{2}-k^{2})f_{\psi}(\omega_{i})
 (\pm k^{2}\mp\omega_{i}^{2}+2p_{0}(k\pm \omega_{i})\mp \delta m^{2})\, ,
\end{align}
where  $\delta m^2\equiv m_\chi^2-m_\phi^2<0$ and the symbol $\Theta_{i}$ imposes  a restriction on  the momentum integration from Eq.~\eqref{eq:SigmachiR}. Integrating the angle via the Dirac $\delta$-function $\delta[(K-P)^{2}-m_{\phi}^{2}]$ in Eq.~\eqref{eq:SigmachiR}, we find that in the timelike region $K^2>0$   the restriction  turns out to be
\begin{align}
\frac{K^{2}+\delta m^{2}}{2(k_{0}+k)}<p_{0}<\frac{K^{2}+\delta m^{2}}{2(k_{0}-k)}\,,\quad  k_{0}-p_{0}>0\, .
\end{align}
Therefore,  $\Theta_{i}$ is given by the Heaviside $\theta$-function with
\begin{align}
	\Theta_{i}=\theta\left[(2p_{0}k)^{2}-(\omega_i^{2}-k^2+\delta m^2 -2p_{0}\omega_i)^{2}\right].
\end{align}

The solutions $\omega_{1,2}$ from the modified 
dispersion relation are shown in Fig.~\ref{fig:dispersion} for $k/T< 1$ and $k/T>1$, respectively. It can been seen that when $k$ becomes larger, the $\omega_1$-mode approaches a dispersion relation $\omega_1\approx k$ while the $\omega_2$-mode approaches a vacuum-like dispersion relation with an asymptotic mass $\sqrt{2}m_\psi(T)$~\cite{Kiessig:2010pr,Kiessig:2011fw,Kiessig:2011ga,Drewes:2013iaa}. 
It allows us to compute Eq.~\eqref{eq:Cchidecay} with the following approximations:
 \begin{align}\label{eq:disperionapprox}
 	\omega_1^2-k^2\approx0\, , \quad \omega_2^2-k^2\approx 2m_\psi^2(T)\, .
 \end{align}
 Note that when the thermal coupling $\kappa$  becomes smaller, the above approximations can already be accurate at lower momenta, as can be seen from  the top of Fig.~\ref{fig:dispersion}. 
With Eq.~\eqref{eq:disperionapprox},  the collision rate of the forbidden decay $C_{\chi,\rm dec}$ reads
\begin{align}\label{eq:Cchidecayapprox}
	C_{\chi,{\rm dec}}\approx\frac{y_{\chi}^{2}}{16\pi^{3}}\int_{m_{\chi}}^{\infty}dp_{0}f_{\chi}^{{\rm eq}}(p_{0})\int_0^\infty dk  \Theta_{2}f_{\psi}(\omega_{2})
	\left(2m_\psi^2(T)+2p_{0}(k-\omega_{2})+ \delta m^{2}\right).
\end{align}

%-----------------------------------------------------------------------------------------------------------------------
\subsubsection{Tree-level amplitude}
%-----------------------------------------------------------------------------------------------------------------------
To see whether we can directly use the vacuum tree-level amplitude to compute the collision rate with the fermion thermal mass put in by hand, let us now calculate the relevant tree-level amplitude.  
%Note that there is a subtlety in computing the amplitude in the center-of-mass frame. Since $\psi$ has a momentum-dependent thermal mass, the dispersion relation~\eqref{eq:fermiondisper} is   not  Lorentz invariant. It points out that the thermal mass varies in different frames we choose. 
%Nevertheless, according to the asymptotic behavior of $\omega_{1,2}$ shown in Fig.~\ref{fig:dispersion}, $\omega_{1,2}$ have an approximate relation $\omega_i^2\approx k^2+m_\psi^2$ at low-momentum regime $k\ll T$, which was   adopted in leptogenesis scenarios~\cite{Giudice:2003jh,Li:2020ner,Li:2021tlv}. On the other hand, it was noticed in Ref.~\cite{Kiessig:2010pr} that   the $\omega_{1}$-mode turns  massless at high-$k$ limit while the $\omega_{2}$-mode  has an approximate relation $\omega_2^2\approx k^2+2m_\psi^2$ with an asymptotic mass $\sqrt{2}m_\psi(T)$ rather than $m_\psi(T)$. 
As can be seen from Fig.~\ref{fig:dispersion}, the $\omega_{1}$-mode quickly turns massless  while the $\omega_{2}$-mode has an asymptotic mass $\sqrt{2}m_\psi (T)$ so that sufficient  momentum space is opened in this mode for the forbidden decay. In the following, we will use  the dispersion relation $\omega^2-k^2=2m_\psi^2(T)$   to calculate the decay rate from the tree-level amplitude.

 The squared amplitude of $\psi\to\chi+\phi$ is given by
\begin{align}
	\sum_{s}|\mathcal{M}|^{2}\approx y_{\chi}^{2}(2\kappa^2 T^2-m_{\phi}^{2})\,,
\end{align}
where the approximation is obtained in the limit $m_{\chi}\ll m_{\phi}$. Note that the squared amplitude for the dispersion relation $\omega^2-k^2=m_\psi^2(T)$ can be  simply obtained by replacing $\sqrt{2}\kappa$ with $\kappa$.

%It should also be noticed that an chirality-flipping amplitude proportional to $m_{\psi}(T)$ is not a physically consistent result since $m_{\psi}(T)$ does not break chirality. Here, we can see that this is not the case. The mass dependence arises from the effective dispersion relation rather than a chirality flip of $\psi$.

The collision rate is given by
\begin{align}\label{eq:Cchidecay2}
C_{\chi,{\rm dec}}	&=\int\frac{d^{3}p_{\psi}}{(2\pi)^{3}2E_{\psi}}f_{\psi}^{{\rm eq}}\int\frac{d^{3}p_{\chi}}{(2\pi)^{3}2E_{\chi}}\frac{d^{3}p_{\phi}}{(2\pi)^{3}2E_{\phi}}(2\pi)^{4}\delta^{4}(P_{\psi}-P_{\chi}-P_{\phi})\sum_{s}|{\cal M}|_{\psi\to\chi\phi}^{2}	
\nonumber \\[0.2cm]
&\approx\frac{y_{\chi}^{2}\kappa^{3}K_{1}(\sqrt{2}\kappa)}{8\sqrt{2}\pi^{3}}\left(1-\frac{m_{\phi}^{2}}{2\kappa^{2}T^{2}}\right)^{2}T^{4}\, ,
\end{align}
where $K_1$ is the modified Bessel function with $K_1(x)\approx 1/x$ for $x<1$. In the last approximation we have used the Boltzmann distribution $f_\psi(E_\psi)=e^{-E_\psi/T}$ and kept the highest scale $m_\phi$ from the dark sector. 

%------------------------------------------------------------------------------------------------------
\begin{figure}[t]
	\centering
	\includegraphics[scale=0.58]{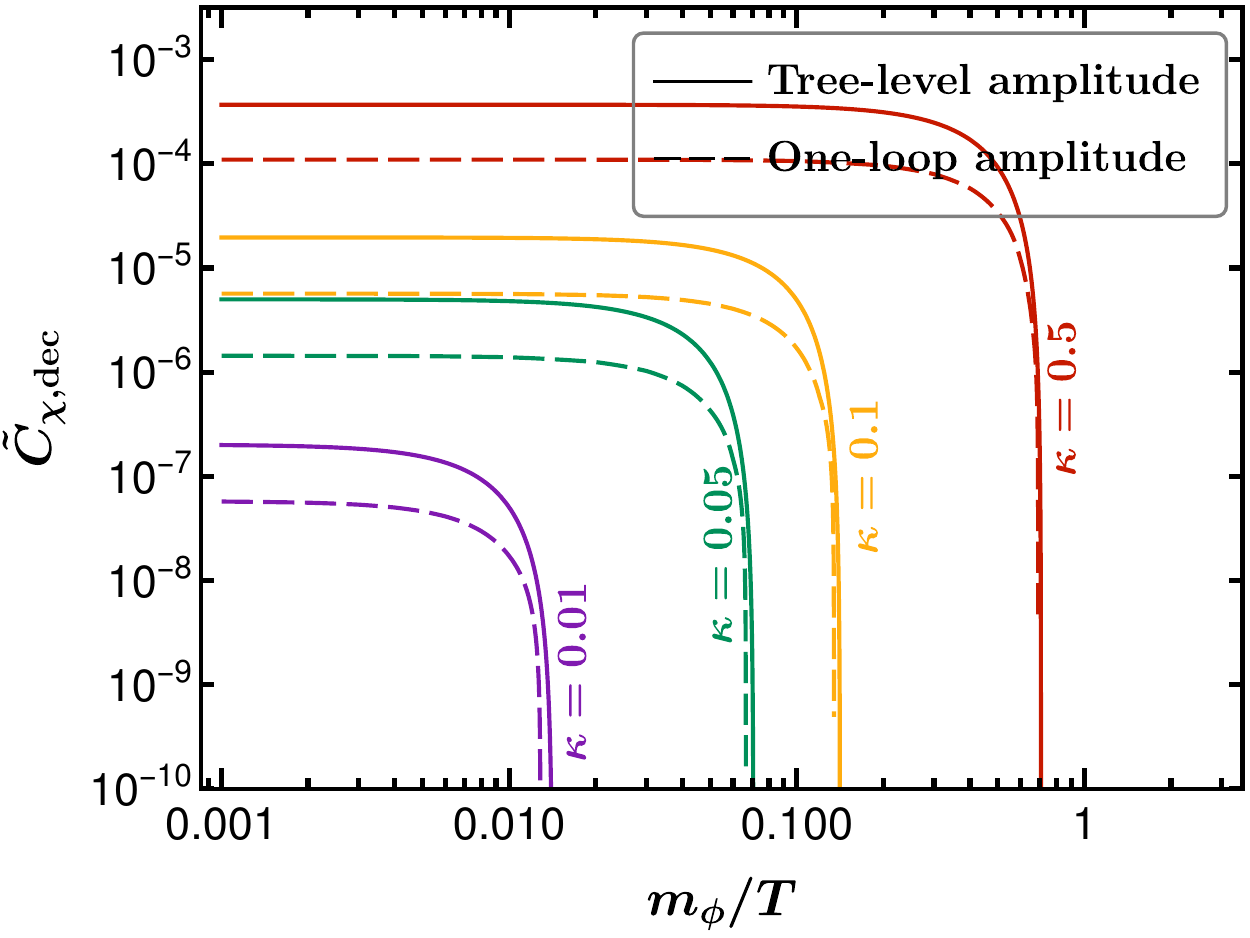}\qquad
	\includegraphics[scale=0.54]{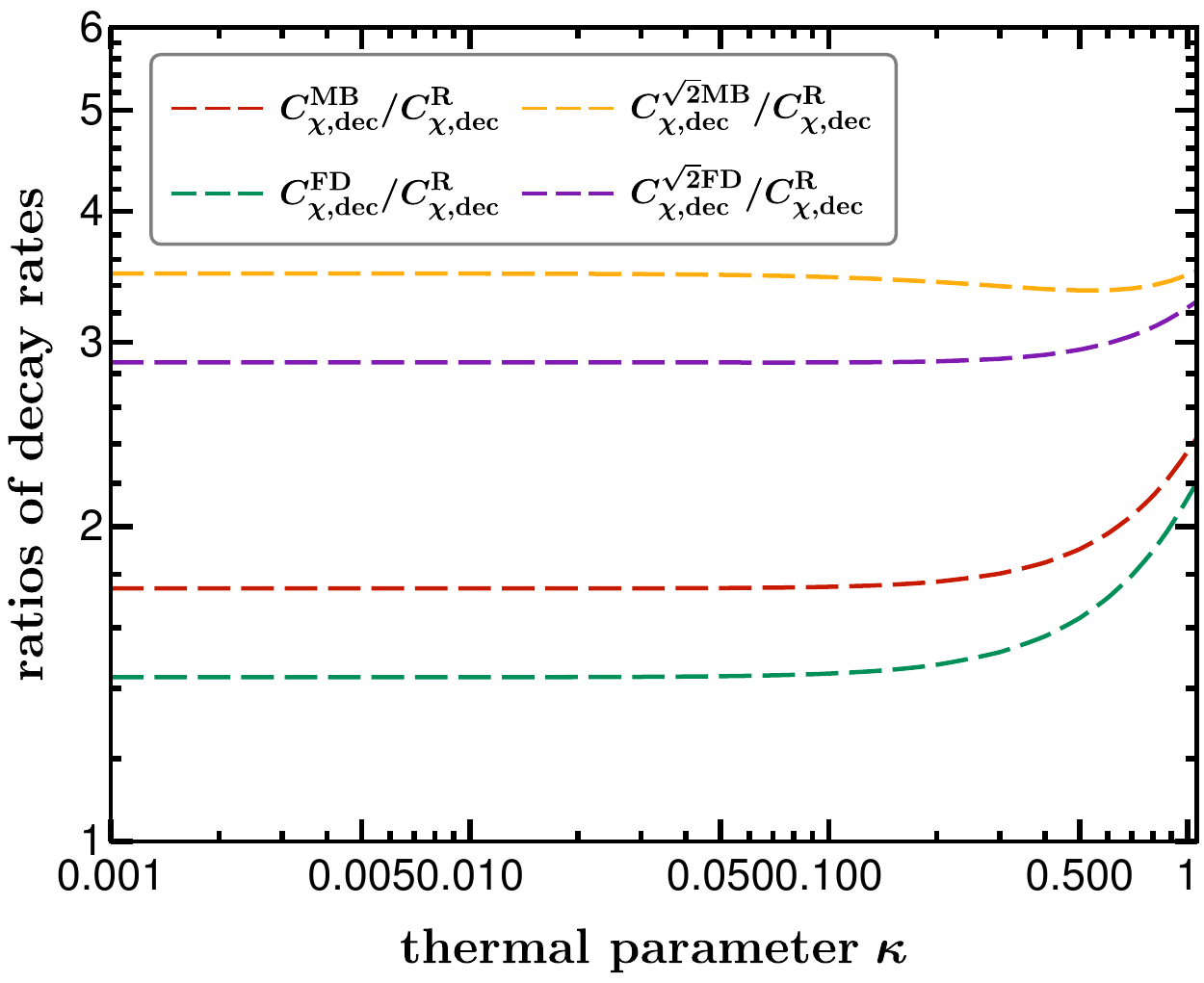}
	\caption{The comparison of forbidden decay rates from the one-loop retarded  and vacuum tree-level amplitudes. Here $\tilde{C}_{\chi,\rm dec}\equiv  y_\chi^{-2}T^{-4}C_{\chi,\rm dec}$. For $\kappa<1$, the rates from the tree-level amplitude  are  overestimated by a factor of $1\textendash4$.
		\label{fig:decaycomparison}
}
	
\end{figure}
%------------------------------------------------------------------------------------------------------

In the left panel of Fig.~\ref{fig:decaycomparison}, we compare the decay rates obtained from Eq.~\eqref{eq:Cchidecayapprox} and Eq.~\eqref{eq:Cchidecay2} with different thermal parameter   $\kappa$. Note that the rates from the two approaches share the same critical temperature 
\begin{align}
T_{c}\approx \frac{m_\phi}{\sqrt{2}\kappa},
\end{align}
after which the decay is kinematically closed.  We can see that the  rate from the tree-level amplitude  with an effective mass $\sqrt{2}m_\psi(T)$ is   overestimated with respect to that from the one-loop retarded amplitude.

In the right panel of Fig.~\ref{fig:decaycomparison}, we also show the ratios of various decay rates by evaluating the vacuum tree-level amplitude with an effective mass $m_\psi(T)$ and taking the full Fermi-Dirac statistics  for $f_\psi^{\rm eq}$. 
Noticeably, a larger discrepancy between the retarded rate $C_{\chi,\rm dec}^{\rm R}$ and the vacuum one appears when the tree-level amplitude is evaluated with the asymptotic mass $\sqrt{2}m_\psi(T)$, as seen from the $C_{\chi,\rm dec}^{\sqrt{2}\rm FD}/C_{\chi,\rm dec}^{\rm R}$ and  $C_{\chi,\rm dec}^{\sqrt{2}\rm MB}/C_{\chi,\rm dec}^{\rm R}$  curves.  Instead, the vacuum rates with the dispersion relation  $\omega^2-k^2=m_\psi^2(T)$ are more compatible with the retarded one.
We found that for $\kappa<0.2$ the ratios become   constants and    reach 
\begin{align}
\frac{C_{\chi,\rm dec}^{\rm FD}}{C_{\chi, \rm dec}^{\rm R}}\simeq 1.4\, , \quad \frac{C_{\chi,\rm dec}^{\rm MB}}{C_{\chi, \rm dec}^{\rm R}}\simeq 1.7\, ,
\end{align}
in which   $C_{\chi,\rm dec}^{\rm FD}$ and  $C_{\chi,\rm dec}^{\rm MB} $ denote the vacuum rates with   the   Fermi-Dirac and Maxwell-Boltzman statistics, respectively, together with the dispersion relation $\omega^2-k^2=m_\psi^2(T)$.  In particular,    a smaller discrepancy can be seen between $C_{\chi,\rm dec}^{\rm FD}$ and $C_{\chi,\rm dec}^{\rm R}$, since    the latter is  also derived from the full Fermi-Dirac statistics. 
%It points out that  the   decay rate from the tree-level amplitude can coincide with that from the   one-loop retarded amplitude within a factor of 2, if $\omega^2-k^2=m_\psi^2(T)$ is put in by hand in the tree-level amplitude.

Since the ratios shown in the right panel of Fig.~\ref{fig:decaycomparison} are predicted via a common thermal parameter $\kappa$, 
%and the ratios become nearly constant when $\kappa<0.2$, 
   the forbidden fermion decay rate can then  be simply obtained from the tree-level amplitude with the approximate dispersion relation $\omega^2-k^2\approx m_\psi^2(T)$ and   rescaling the latter by a constant read from the figure. It enables us to   obtain a rather precise forbidden fermion decay rate within the simple tree-level approach  by $\kappa$-dependent constant rescaling. The results shown in this section not only confirm that using $\omega^2-k^2\approx m_\psi^2(T)$ in the tree-level amplitude for forbidden decay is a good approximation~\cite{Giudice:2003jh,Kiessig:2010pr}, but also provide quantitative differences characterized only by the thermal parameter.

%The underlying reason for the difference between the two approaches dates back to the usage of effective mass in the tree-level amplitude. While there exist momentum space for $\omega_{i}^{2}-k^{2}=2m_{\psi}^{2}(T)$ or $\omega_{i}^{2}-k^{2}=m_{\psi}^{2}(T)$ to a very good approximation. However, for smaller $\kappa$, the Lambert W-function blows up quickly, leading to $\omega_{i}=k$ and the momentum space for the approximate dispersion relation   becomes smaller, thereby leading to a smaller rate under the $k$-integration. Such shrink space is not followed by the analytic decay rate from the tree-level amplitude in which the approximate dispersion relation is still adopted  in the  high-$k$ limit, therefore resulting in  the overestimate.

%-----------------------------------------------------------------------------------------------------------------------
\section{Scattering}\label{sec:scattering}
%-----------------------------------------------------------------------------------------------------------------------

%-----------------------------------------------------------------------------------------------------------------------
\subsection{Double counting and resonant enhancement}
%-----------------------------------------------------------------------------------------------------------------------
The scattering rate directly calculated from Fig.~\ref{fg:SigmaRchi}  is much more involved. The   imaginary parts $\text{Im}\Delta_{\pm}$ appear   both in the numerator and   denominator of the off-shell spectral density $\rho_{\psi,{\rm off}}$, making the  final  three-dimensional integration ($dp dk_0 dk$) difficult even with a numerical approach.   For most situations, the thermal corrections to the scattering processes are significant only when there are IR singularities or resonance. For example, the IR singularity is known in  neutrino and electron  chirality-flipping processes at finite temperatures~\cite{Elmfors:1997tt,Ayala:1999xn,Boyarsky:2020cyk,Li:2022dkc}, and the resonant effect from thermal corrections is also known in neutrino oscillations at finite temperature and density~\cite{Wolfenstein:1977ue,Notzold:1987ik}.

In dealing with  the IR singularity or resonance,  we can also use a   more convenient approach  in which the cross section is calculated from a tree-level diagram  with a resummed   mediator propagator~\cite{Blaizot:1999xk,Blaizot:2001nr,Arnold:2002zm}. 
%This corresponds to an effective kinetic equation at finite temperatures, and the result from this effective approach has been shown in the  neutrino  chirality-flipping process to be consistent with that from the full thermal calculation~\cite{Li:2022dkc}.
When applying the effective approach, however, we should take care of the double-counting issue. There are in general two  methods to  remove the double counting.  
When the full thermal width of the mediator propagator is unknown, it is   convenient to subtract the on-shell point directly from the cross section, and then calculate the forbidden decay rate separately. On the other hand, if the thermal width is known in a given   model,  a modified Breit-Wigner approximation can be applied to do the subtraction~\cite{Belanger:2018ccd,Bringmann:2021sth}, where the decay is automatically included in the cross section. 

Nevertheless,    the  double-counting  issue  depends on the existence of the resonance, which requires a careful inspection under the perturbative HTL resummation.
In the following, let us concentrate on the $s$-channel double counting
%\footnote{In some cases, there could also be $t/u$-channel double counting at finite temperatures. See e.g. Refs.~\cite{Giudice:2003jh,Grzadkowski:2021kgi}} 
and on the   hard   particle scattering with  incoming momenta $p_{\rm hard}\sim \mathcal{O}(T)$.  Generically,   hard scattering suffices to be responsible for the nonthermal DM production from thermal particles, since   the thermally averaged collision rate $\langle\sigma v\rangle n$ is proportional to the particle-number densities of incoming thermal particles, which are expected to be dominated in the hard-momentum regime:
\begin{align}
	n_{\rm soft}\propto \int_{0}^{p_{\rm soft}} d^3 p f^{\rm eq}(p)\sim p_{\rm soft}^3,\quad 	n_{\rm hard}\propto \int_{p_{\rm soft}}^{\infty} d^3 p f^{\rm eq}(p)\sim T^3\gg p_{\rm soft}^3\, ,
\end{align}
with   $p_{\rm soft}\sim \mathcal{O}(y_\psi T)$.

At leading order, the mediator is resummed while the external particles are treated effectively massless. At this order, it is usually expected to have an $s$-channel resonance when the momentum transfer is near the scale of the effective mediator mass. However, when we go beyond the leading order, the external particles are   resummed, which also carry effective masses from the plasma. If the  thermal masses from the external particles are larger than from  the mediator, the resonance expected  at leading order  would be erased.  This is interpreted as the fact that the inverse decay $X+Y\to Z$ is always kinematically forbidden at all temperatures. This is particularly the case when the mediator is a fermion and the incoming particles contain a scalar  boson. For instance, the resummed scalar $\varphi$ has a thermal correction parameter $\kappa=y_\psi/\sqrt{12}$~\cite{Li:2022rde} from the $\psi$-$\eta$ loop, which is larger than the value given in Eq.~\eqref{eq:thermalmass}. 
%On the other hand, similar to  the relativistic QED plasma~\cite{Elmfors:1997tt,Ayala:1999xn,Li:2022dkc}, a resummed vector boson from the $\psi$-loop has a    thermal correction parameter $\kappa=g_\psi/\sqrt{6}$, which is also larger than the value given in Eq.~\eqref{eq:thermalmass}. 

The above conclusion differs from  two fermion scattering mediated   by a thermal scalar. As seen from Fig.~\ref{fig:dispersion}, there is a nearly massless  state for a resummed fermion so that   the initial fermions can have an approximate dispersion relation $\omega_i^2-k^2\approx 0$  while the resummed scalar mediator  carries a large thermal mass. When  the momentum transfer is at the order of the scalar thermal mass, there is in principle an on-shell crossing and including the resummed scalar mediator in the fermion-pair scattering can enhance the scattering rate by a factor of $\mathcal{O}(1)$~\cite{Li:2022rde}.

%^It should also be mentioned  that  even with soft-momentum transfer $s\simeq p_{\rm soft}^2$, in which higher-order HTL corrections (resummed vertex and resummed initial particles) should be included, there is no resonance when the initial particles contain scalar/vector boson. The resummed initial fermion is expected to have a similar dispersion relation to the fermion mediator. It implies that, at soft scattering regime, the  total thermal mass  effect from initial fermion and scalar/vector boson is  larger than the fermion mediator, so there is no on-shell crossing happening  in the fermion mediator.

Since  in current scenario   the initial particles contain a  fermion and a scalar  boson, it is not necessary to use  the resummed fermion mediator and the scattering rate from a vacuum computation suffices to describe the DM production to a good approximation.

%-----------------------------------------------------------------------------------------------------------------------
\subsection{Tree-level scattering amplitude without thermal correction}
%-----------------------------------------------------------------------------------------------------------------------
The general $2\to2$ scattering rate for the DM production is given by
\begin{align}
	C_{12\to\chi\phi}&=\int\frac{d^{3}p_{1}}{(2\pi)^{3}2E_{1}}\frac{d^{3}p_{2}}{(2\pi)^{3}2E_{2}}\frac{d^{3}p_{\chi}}{(2\pi)^{3}2E_{\chi}}\frac{d^{3}p_{\phi}}{(2\pi)^{3}2E_{\phi}}f_{1}f_{2}|{\cal M}|_{12\to\chi\phi}^{2}(2\pi)^{4}\delta^{4}\, ,
\end{align}
where $\delta^{4}\equiv\delta^{4}(P_{1}+P_{2}-P_{\chi}-P_{\phi})$ and $|{\cal M}|_{12\to\chi\phi}^{2}$ is the squared amplitude with spin sum but without spin average. The Pauli blocking and Bose enhancement from the nonthermal DM sector are neglected.

%------------------------------------------------------------------------------------------------------
\begin{figure}[t]
	\centering
	\includegraphics[scale=0.7]{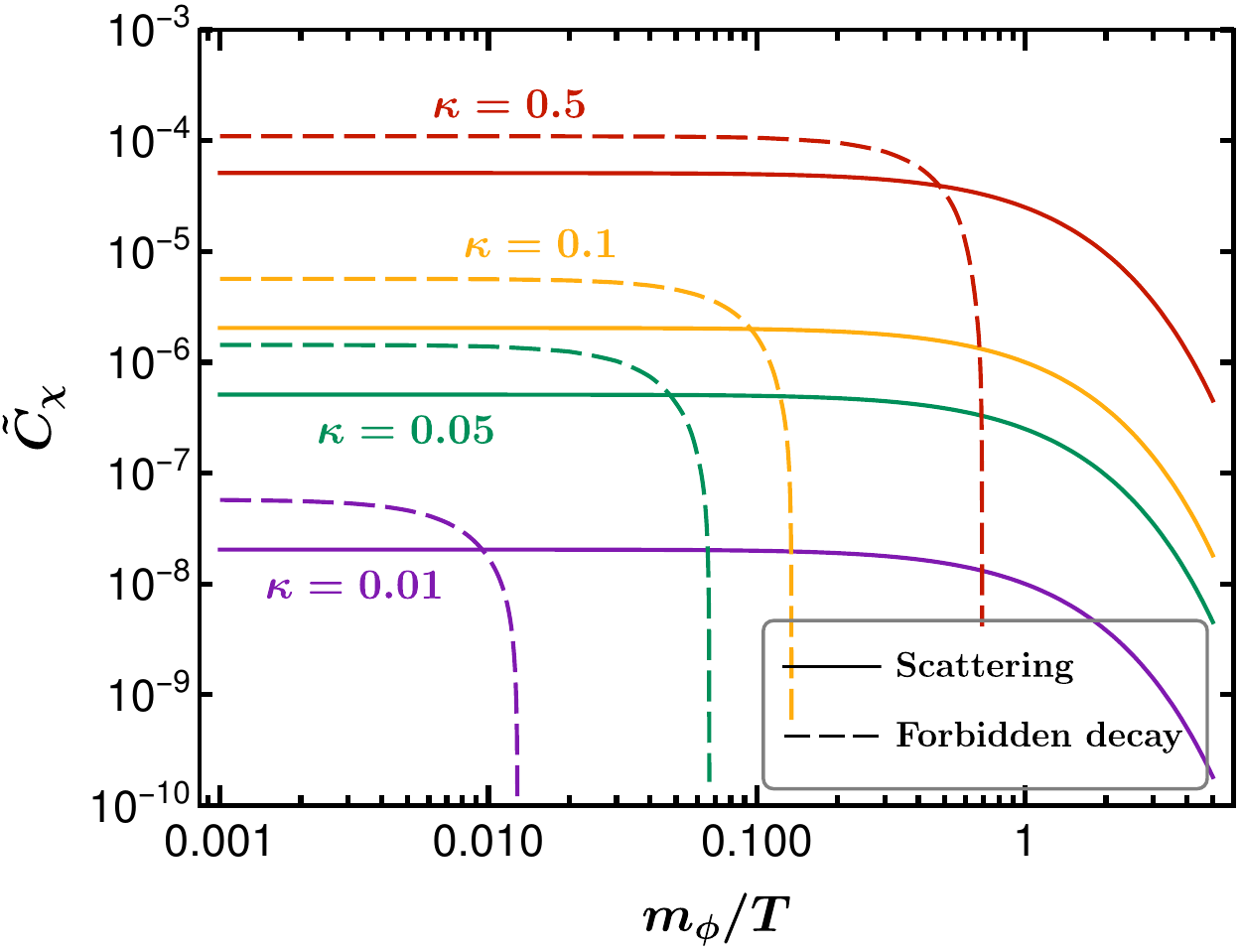}
	\caption{A comparison between the forbidden decay and scattering rates for different thermal parameter $\kappa$. Here $\tilde{C}_\chi\equiv y_\chi^{-2}T^{-4} C_\chi$.
		\label{fig:decayVSscat}
	}
	
\end{figure}
%------------------------------------------------------------------------------------------------------

For Yukawa interaction, the scattering is $\eta+\varphi \to\chi+\phi$. The squared amplitude  is given by
\begin{align}
\sum_{s}|\mathcal{M}|_{\varphi \eta\to\chi\phi}^{2}	
\approx\frac{y_{\chi}^{2}y_{\psi}^{2}}{2}(1-\frac{m_{\phi}^{2}}{s})(1+\cos\theta)\, ,
\end{align}
where we have only kept  the highest mass scale from  $m_\phi$ and $\theta$ is the angle between the spatial momenta of the incoming and outgoing particles in the center-of-mass frame.
Following the conventional phase-space reduction~\cite{Gondolo:1990dk},  we obtain the collision rate
\begin{align}\label{eq:Cscat}
	C_{\varphi\eta\to\chi\phi}	=\frac{T}{32\pi^{4}}\int_{m_{\phi}^{2}}^{\infty}ds\sigma_{\varphi\eta\to\chi\phi}s^{3/2}K_{1}(\sqrt{s}/T)\, ,
\end{align}
where the  cross section without spin average is given by
\begin{align}
\sigma_{\varphi\eta\to\chi\phi}=\frac{y_{\chi}^{2}y_{\psi}^{2}}{32\pi s}\left(1-\frac{m_{\phi}^{2}}{s}\right)^{2}\,.
\end{align}
In  the high-temperature limit $T\gg m_\phi$, the collision rate reduces to 
\begin{align}\label{eq:scatlimit}
	C_{\varphi\eta\to\chi\phi}\approx\frac{y_{\chi}^{2}y_{\psi}^{2}}{256\pi^{5}}T^{4}\, .
\end{align}

In Fig.~\ref{fig:decayVSscat}, we show the rates from the   forbidden decay and scattering channels.  In general,  $C_{\chi,\rm dec}$ is larger than $C_{\chi,\rm scat}$ when $T>T_c$. Nevertheless,    the duration of the forbidden decay     is determined by the critical temperature $T_c$, while the scattering $\eta+\varphi \to\chi+\phi$ is sufficiently closed only after  the freeze-in temperature  $T\sim m_\phi> T_c$. It makes the scattering contribution to the final DM relic density generically larger than the forbidden decay, as we shall discuss below.

%-----------------------------------------------------------------------------------------------------------------------
\section{DM relic density}\label{sec:relicdensity}
%-----------------------------------------------------------------------------------------------------------------------

There are in principle two possibilities for DM relic density. If the scalar $\phi$ is unstable, it can decay to $\chi$ at late times after the dark sector freezes in. Consider first the situation where $\phi$ has been depleted away.  $\chi$ is the DM candidate and the relic density is   given by
\begin{align}\label{eq:DMrelic1}
\Omega_{{\rm DM}}h^{2}=\frac{(Y_{\chi}^{{\rm I}}+Y_{\chi}^{{\rm II}})s_{0}m_{\chi}}{\rho_{c}/h^{2}}\,.
\end{align}
where $Y_{\chi}^{{\rm I}}\equiv n_{\chi}^{\rm I}/s_{\rm SM}$ is the yield produced by forbidden decay and scattering while $Y_{\chi}^{{\rm II}}$ is the yield produced by scalar decay $\phi\to\psi+\chi$ at late times.  $s_{\rm SM}=2\pi^2 g_s T^3/45$ is the SM entropy density with $g_s$ the  effective number of relativistic degrees of freedom. The current value of entropy density is given by $s_0=2891.2~\text{cm}^{-3}$ and the current critical energy density $\rho_c$ is given by $\rho_c=1.05\times 10^{-5}~h^2\cdot \text{GeV}\cdot \text{cm}^{-3}$~\cite{ParticleDataGroup:2022pth}.

The Boltzmann equation for $Y_{\chi}^{{\rm I}}$ is given by
\begin{align}\label{eq:YBoltzmann}
	Y^{\rm I}_\chi=\int_{T_c}^\infty\frac{2C_{\chi, \rm dec}}{s_{\rm SM} H T}dT+\int_{0}^\infty\frac{2C_{\chi, \rm scat}}{s_{\rm SM} H T}dT\, ,
\end{align}
where the factor of 2 accounts for the $CP$-conjugated production so that $Y_\chi$ is the sum of $\chi+\bar\chi$.  The forbidden decay ends at $T=T_c$ while the scattering basically ends at $T=\mathcal{O}(m_\phi)$ as the freeze-in temperature is determined by the highest scale in the dark sector. In the  second term of  Eq.~\eqref{eq:YBoltzmann}, we use $T=0$ as the lower integration limit, which does not cause significant difference after $T$ drops below $m_\phi/5$.
Since  both $\chi+\bar\chi$ and $\phi$ are produced with the same amount from the forbidden decay and scattering,  we have $Y_{\chi}^{{\rm I}}=Y_{\phi}^{{\rm I}}$. Further given that the amount of  $\chi+\bar\chi$ in    late-time production is inherited from $Y_{\phi}^{{\rm I}}$, we have $Y_{\chi}^{{\rm II}}=Y_{\phi}^{{\rm I}}$.

Consider the second possibility where $\phi$ is sufficiently long-lived so that it has a lifetime comparable with or longer than the age of the observed universe.   The DM relic density in this case consists of $\phi$ and $\chi$,  which is given by
\begin{align}\label{eq:DMrelic2}
\Omega_{{\rm DM}}h^{2}=\frac{s_{0}}{\rho_{c}/h^{2}}(Y_{\chi}^{{\rm I}}m_{\chi}+Y_{\phi}^{{\rm I}}m_{\phi})\,.
\end{align}

To see the relative effect of the forbidden decay and the scattering channel, we   estimate the ratio $Y_{\chi,\rm scat}/Y_{\chi,\rm dec}$, which reads:
\begin{align}\label{eq:Yratio}
	\frac{Y_{\chi,\rm scat}}{Y_{\chi,\rm dec}}\approx \frac{\int_{0}^{x_{\phi,\rm fi}} \tilde{C}_{\chi,\rm scat}dx_\phi}{\int_{0}^{\sqrt{2}\kappa} \tilde{C}_{\chi,\rm dec} dx_\phi}\, ,
\end{align}
where $x_\phi\equiv m_\phi/T$ with $x_{\phi,\rm fi}$ corresponding to the freeze-in temperature. The evolution of $\tilde{C}_{\chi,\rm dec}$ and $\tilde{C}_{\chi,\rm scat}$ can be seen  in Fig.~\ref{fig:decayVSscat}.  
%To obtain an analytic behavior of Eq.~\eqref{eq:Yratio}, we use the high-temperature limit of $C_{\chi,\rm dec}$ in Eq.~\eqref{eq:Cchidecay2}, 
%\begin{align}\label{eq:Cchidecay2}
%	\tilde{C}_{\chi,{\rm dec}}\approx \frac{\kappa^{3}K_{1}(\sqrt{2}\kappa)}{8\sqrt{2}\pi^{3}}\, ,
%\end{align}
%and the high-temperature limit of $C_{\chi,\rm scat}$ from Eq.~\eqref{eq:scatlimit}. 
Simply taking  $\tilde{C}_{\chi,\rm dec}$ and $\tilde{C}_{\chi,\rm scat}$ as constants,  we obtain $Y_{\chi,\rm scat}/Y_{\chi,\rm dec}\propto 1/\kappa$. 
It points out that the DM relic density from the forbidden decay basically carries an additional  power of $\kappa$ higher than  from the scattering channel, even though both the decay and scattering rates share the same order of $\kappa$ (see Eqs.~\eqref{eq:Cchidecay2} and~\eqref{eq:Cscat}), as also found in Refs.~\cite{Darme:2019wpd,Li:2022rde} in the case of    forbidden scalar decay.  
%In general, besides having  a prefactor $\mathcal{O}(\kappa^3)$, $Y_{\chi,\rm dec}$ also depends on $\kappa$ via the momentum integration in $C_{\chi,\rm dec}$. As $\kappa$ decreases, the momentum space for the forbidden decay, i.e., the regions for $\omega_i^2-k^2>0$, becomes smaller. Thus for much smaller $\kappa$, the ratio  $Y_{\chi,\rm dec}/Y_{\chi,\rm scat}$ is more suppressed than the expectation  $Y_{\chi,\rm dec}/Y_{\chi,\rm scat}\simeq \mathcal{O}(\kappa)$ which occurs in the scalar mediator case.

\begin{figure}[t]
	\centering
	\includegraphics[width=0.47\textwidth]{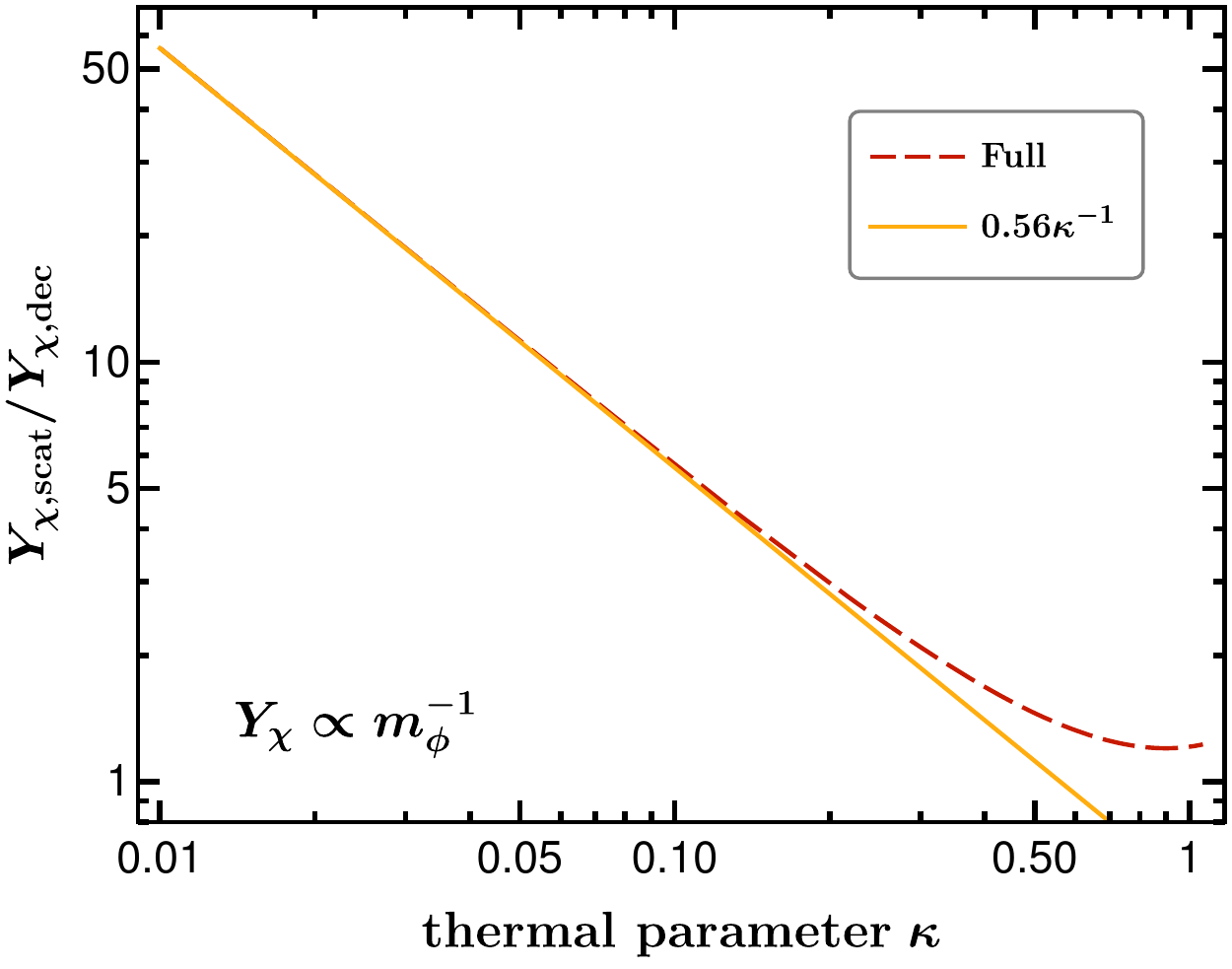}\quad
	\includegraphics[width=0.49\textwidth]{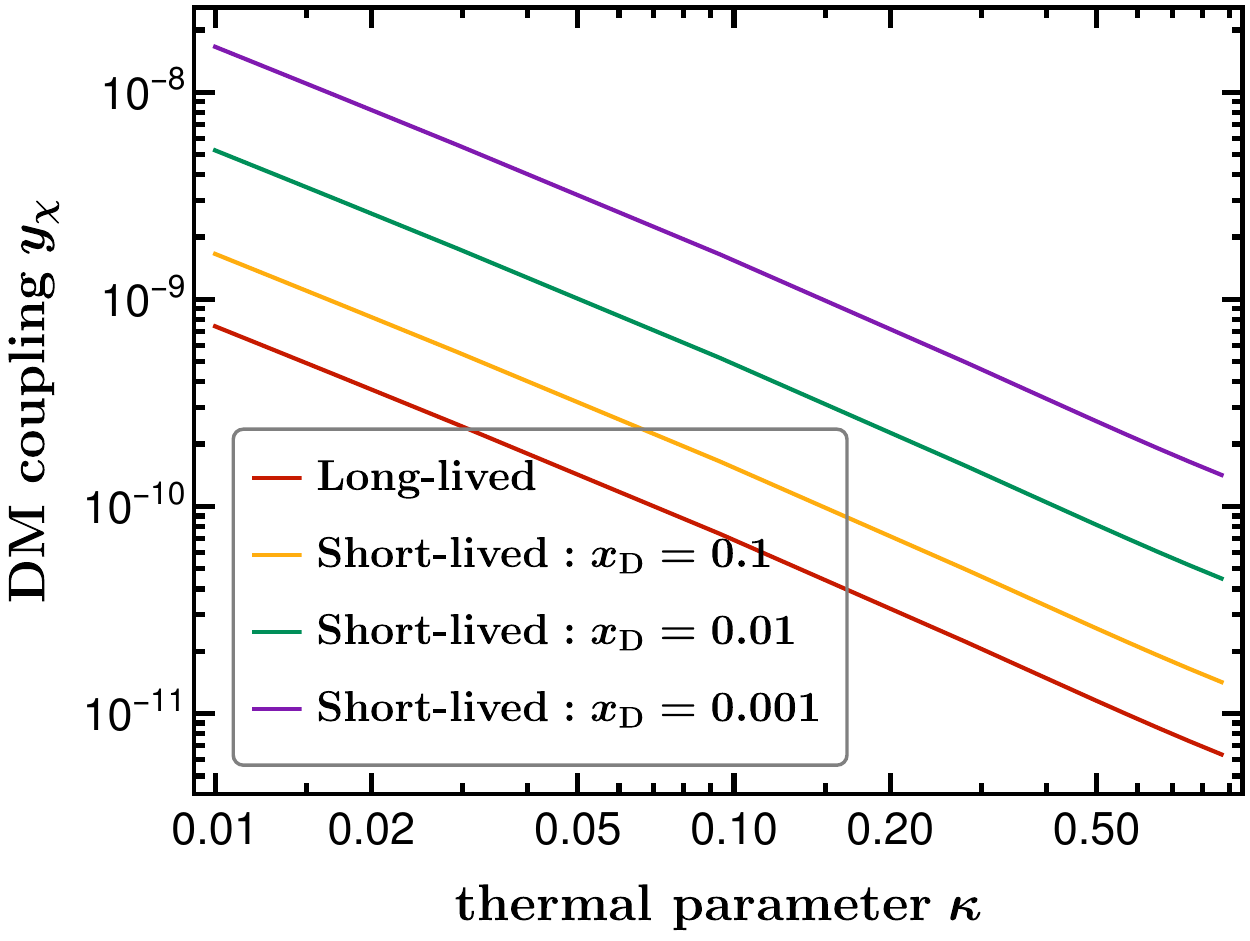}
	\caption{Left: A comparison of DM relic densities from the forbidden decay and scattering channels. Right: The correlation between the DM coupling $y_\chi$ and the thermal parameter $\kappa$ for  the observed DM relic density.  Here $x_D\equiv m_\chi/m_\phi$.
		\label{fig:decayVSscatRatio}
	}
	
\end{figure}

The behavior of Eq.~\eqref{eq:Yratio} is shown in  the left panel of  Fig.~\ref{fig:decayVSscatRatio} as a function of the thermal parameter $\kappa$. Note that  only the highest scale $m_\phi$ is kept in the yield so that both  $Y_{\chi,\rm dec}$ and $Y_{\chi,\rm scat}$ are proportional to the inverse scalar mass, as expected from the IR freeze-in mechanism. 
We can see from the   left panel of Fig.~\ref{fig:decayVSscatRatio} that for the fermion mediator  the forbidden decay can only   be neglected   for a  very small $\kappa$. For a large $\kappa$, the contributions from the forbidden decay and the scattering could be comparable. 
For instance,  about $41\%$ of the DM relic density from Eq.~\eqref{eq:DMrelic1} comes from the forbidden decay if  $\kappa=0.5$, while about $8\%$ of the DM relic density is obtained from the forbidden decay if   $\kappa=0.05$.  As  already mentioned in  Sec.~\ref{sec:spectraldensity}, we have taken a conservative limit for a Yukawa coupling $y_\psi<1$, and the upper limit of $\kappa$ depends on the gauge degeneracy of thermal particles $\eta,\varphi$ and on possible flavored Yukawa couplings.  These effects could further enhance the  decay contribution, which is the subject of  Sec.~\ref{sec:UVmodels}. 

An interesting feature from such a comparison is that we can estimate the effect of  the forbidden
 decay by rescaling the scattering rate, since the ratio given in Eq.~\eqref{eq:Yratio}   basically depends on the thermal coupling $\kappa$, or the interaction coupling $y_\psi$. Once the thermal interaction of the fermion mediator is known, we can calculate the scattering rate and simply rescale it  by a $\kappa$-  or $y_\psi$-dependent factor to obtain  the forbidden decay. As shown in the left panel  of Fig.~\ref{fig:decayVSscatRatio}, when $\kappa\lesssim 0.2$, the ratio is approximately given by $0.56/\kappa$ and the total DM relic density given in Eq.~\eqref{eq:DMrelic1} can then be estimated by
 \begin{align}
\Omega_{{\rm DM}}h^{2}\approx 2\frac{s_{0}m_{\chi}}{\rho_{c}/h^{2}}(1+1.79\kappa)Y_{\chi, \rm scat}\, ,
 \end{align}
where $Y_{\chi,\rm scat}$ comes from the second term in Eq.~\eqref{eq:YBoltzmann}.
%------------------------------------------------------------------------------------------------------
%------------------------------------------------------------------------------------------------------

In the right panel of Fig.~\ref{fig:decayVSscatRatio}, we plot the correlation between the DM coupling $y_\chi$ and the thermal parameter $\kappa$ by fitting the observed DM relic density $\Omega_{\rm DM }h^2=0.12$~\cite{Planck:2018vyg}.  The long-lived line corresponds to the second possibility from Eq.~\eqref{eq:DMrelic2}, where we have neglected the contribution from the light $\chi$. In this approximation, the DM relic density is independent of $m_\phi$ since $Y_{\phi}^{\rm I}\propto m_\phi^{-1}$.  However, the DM relic density from Eq.~\eqref{eq:DMrelic2} requires that the scalar should have a lifetime longer than the age of the universe, which is translated into an upper limit of the DM coupling $y_\chi\lesssim 10^{-20} (m_\phi/\text{GeV})^{-1/2}$. Therefore, we can conclude from the right panel of Fig.~\ref{fig:decayVSscatRatio} that for a dark scalar heavier than 1 GeV,  the   DM relic density results from the lighter fermion $\chi$. For instance,   with $y_\chi\simeq 10^{-11}$ and $m_\phi\simeq 10$~GeV, the scalar lifetime is around $\tau_\phi\simeq 0.03$~s. Thus the unstable heavy scalar has decayed away well before the big bang nucleosynthesis epoch.

For the short-lived case from Eq.~\eqref{eq:DMrelic1}, the DM relic density depends on $y_\chi, \kappa$ and the mass ratio in the dark sector $x_{\rm D}\equiv m_\chi/m_\phi$. We show in the right panel of Fig.~\ref{fig:decayVSscatRatio} for three representative values $x_{\rm D}=0.1,0.01,0.001$.  We can see that when the mass ratio $x_{\rm D}$ and the thermal parameter $\kappa$ decrease, a larger DM coupling $y_\chi$ is required to match the relic density. However, a large DM coupling could make the dark sector thermalized. To check this, recall that
the nonthermal condition, which requires that the thermally averaged scattering rate should be smaller than the Hubble parameter at the freeze-in temperature, is given by
\begin{align}
	%\langle \sigma v\rangle n_\chi^{\rm eq}=
	\frac{C_{\chi,\rm scat}}{n_{\chi}^{\rm eq}}<H \, ,
\end{align}
where $n_\chi^{\rm eq}\approx 0.09 T^3$ denotes the thermal particle-number density  of $\chi$. The above condition can be  translated into an upper limit of the   DM coupling $y_\chi\lesssim \mathcal{O}(10^{-4})$.   Therefore,  for the thermal parameter $\kappa$   and the   mass ratio   $x_D$  shown in the right panel of Fig.~\ref{fig:decayVSscatRatio}, the dark sector is indeed far from thermal equilibrium.

%%------------------------------------------------------------------------------------------------------
%\begin{figure}[t]
%	\centering
%	\includegraphics[scale=0.5]{Figs/CoupCorre1.pdf}\quad
%		\includegraphics[scale=0.5]{Figs/CoupCorre2.pdf}
%	\caption{Comparison between decay and scattering rates.
%		\label{fig:DMfit}
%	}
%	
%\end{figure}
%%------------------------------------------------------------------------------------------------------

%Finally it is noteworthy that  we only consider a relatively large thermal coupling $\kappa$. 
When $\kappa$ is much smaller but still able to keep the fermion mediator in thermal equilibrium, the scattering  channel for the DM production can also come from the mediator scattering/annihilation, e.g., $\psi+\bar\psi\to \chi+\bar\chi$ mediated by the scalar $\phi$ and $\psi+\bar\psi\to \phi+\phi$ mediated by $\chi$, both of which are not included in previous calculations since we are concerned with a relatively large $\kappa$.  These scattering channels have rates  at $\mathcal{O}(y_\chi^4)$ and could be comparable with the thermal particle scattering $\sim\mathcal{O}(y_\chi^2y_\psi^2)$  if $y_\chi\sim y_\psi$.
For example, when the fermion mediator $\psi$ is a GeV-scale right-handed neutrino in the type-I seesaw framework, the scattering  $\psi+\bar\psi\to \chi+\bar\chi$ that can generate the observed DM relic density predicts a nonthermal DM coupling $y_\chi\sim \mathcal{O}(10^{-6})$ while the coupling for a  GeV-scale right-handed neutrino to keep in thermal equilibrium via neutrino oscillation is required to be  $y_\psi> \mathcal{O}(10^{-8})$~\cite{Dolgov:2003sg,Li:2022bpp}.  Therefore, for a much smaller thermal parameter $\kappa$,  the DM production from the mediator scattering/annihilation could be significant.  A large   thermal parameter $\kappa$, on the other hand,  is usually more favorable as the strong connection between the SM and the fermion mediator enables us to have more opportunities of  DM detection via the very fermion messenger, and is widely predicted in  specific scenarios that can explain  experimental anomalies, as to be discussed in the following.

%-----------------------------------------------------------------------------------------------------------------------
\section{Application to different thermal Yukawa interactions}\label{sec:UVmodels}
%-----------------------------------------------------------------------------------------------------------------------
We have considered a typical example  in Sec.~\ref{sec:simplemodel} where  both the DM-mediator and SM-mediator connections are realized by Yukawa portal interaction.  In this section, we shall discuss some specific models  to which previous calculations can be applied.  The aim of this section is to calculate the DM relic density,  following   previous sections which  combine the forbidden fermion decay and the associated scattering in terms of the thermal parameter $\kappa$. We specify some typical thermal interactions  with different gauge representations for the thermal particles, and discuss the significance of forbidden fermion decay that could be readily overlooked in the scenarios of light fermion mediators. 

It should be mentioned that the   observational signatures of these specific scenarios depend not only on the thermal coupling, but also on the masses of thermal particles and  the mediator. However, the analyses presented in previous sections only assume  that the relevant thermal species are lighter than the heaviest scale in the dark sector. Given that the scale of the dark sector is not known a priori, there is no definite mass limit that can be inferred from the interplay between the mediator and the DM. On the other hand,  we can infer from the right panel of Fig.~\ref{fig:decayVSscatRatio} that   increasing the thermal coupling $\kappa$ and the  mass hierarchy $x_{\rm D}$ in the dark sector can open up the DM parameter space towards smaller values, which could help to evade   severe observational constraints  whenever the detection of DM via the light fermion mediator is concerned.  In doing so, i.e., increasing the thermal parameter, the forbidden decay  cannot be overlooked. In the following analysis, we commonly assume that there is only one fermion mediator that connects the dark sector to the SM thermal plasma. 

%-----------------------------------------------------------------------------------------------------------------------
\subsection{Right-handed fermion mediator}
%-----------------------------------------------------------------------------------------------------------------------
Presumably, the most known example is the Majorana neutrino portal DM~\cite{Falkowski:2009yz,Gonzalez-Macias:2016vxy,Batell:2017rol, Bandyopadhyay:2018qcv,Becker:2018rve, Folgado:2018qlv,Bandyopadhyay:2020qpn,Biswas:2021kio,Coy:2021sse,Barman:2022scg,Li:2022bpp}, but $\psi_R$ can also be identified as the right-handed Dirac  counterpart of the SM left-handed neutrinos. Both the Majorana and Dirac neutrino mediators   allow a dark sector to be produced via the freeze-in mechanism, as long as $\psi_R$ does not have strong gauge interactions. A noticeable difference between the Majorana and Dirac portals is that the latter naturally predicts a very light fermion mediator with mass readily well below the dark scale.

For right-handed neutrino mediators,  the   left-handed fermion in~\eqref{eq:Yukawathermal} is identified as the SM lepton doublet, while the scalar can either be  the SM Higgs  doublet  or   a new Higgs doublet. In the former case,  a light  right-handed Majorana neutrino with   small Yukawa couplings can already be in thermal equilibrium via fast neutrino oscillation~\cite{Dolgov:2003sg,Li:2022bpp}. So if the active-sterile neutrino mixing is small, the thermal corrections to the Majorana neutrino would be suppressed.  Consequently, the duration of the  forbidden decay channel would be quite short and the scattering becomes the dominant channel to generate the DM relic density. Certainly, large Yukawa couplings are still allowed for Majorana neutrinos, in particular, if they couple to a new Higgs doublet.  In this latter case, an  $\mathcal{O}(1)$  Yukawa coupling between the right-handed neutrino and the new Higgs doublet was interesting, as   considered in  Dirac neutrino mass origin~\cite{Gabriel:2006ns,Davidson:2009ha,Li:2022yna} and in explanations of  flavor anomalies observed at low-energy experiments~\cite{Li:2018rax,Crivellin:2019dun,DelleRose:2019ukt,Duan:2021whx}. 

With large Yukawa interactions, the forbidden right-handed neutrino decay cannot be neglected.  Applying the calculation in Sec.~\ref{sec:forbiddendecay} , we can readily obtain the thermal mass for the right-handed neutrino,
\begin{align}\label{eq:thermalmass2}
	m_\psi^2(T)=\frac{y_\psi^2}{8}T^2\,,
\end{align}
and hence $\kappa=y_\psi/\sqrt{8}$. 
Note that the different thermal mass in Eq.~\eqref{eq:thermalmass2} does not modify the ratio  $Y_{\chi,\rm scat}/Y_{\chi,\rm dec}$ in terms of a free $\kappa$. To see this, recall that the calculation of forbidden decay is given in terms of $m_\psi\equiv \kappa T$. From Eq.~\eqref{eq:thermalmass} to Eq.~\eqref{eq:thermalmass2}, we have $m_\psi\equiv \kappa T\to \tilde{\kappa} T$, with $\tilde \kappa=\sqrt{2}\kappa$.  On the other hand, the scattering cross section is now enhanced by a factor of 2 due to the gauge degeneracy, so the yield from Eq.~\eqref{eq:thermalmass} to Eq.~\eqref{eq:thermalmass2} is changed as $Y_{\chi,\rm scat}\propto y_\psi^2=(4\kappa)^2\to 2y_\psi^2=(4\tilde{\kappa})^2$. Therefore, for a  thermal mass different from Eq.~\eqref{eq:thermalmass}, the updated ratio $Y_{\chi,\rm scat}/Y_{\chi,\rm dec}$ can still be   given by a free $\kappa$ but with an enhanced maximum. For example.   the thermal coupling in Eq.~\eqref{eq:thermalmass}   indicates $0.025<\kappa<0.25$ under the condition~\eqref{eq:weakcoup}, and it is  enhanced to be   $0.035<\kappa<0.35$ in Eq.~\eqref{eq:thermalmass2}.

From the left panel of Fig.~\ref{fig:decayVSscatRatio}, we can now obtain the portion from the forbidden decay channel to the DM relic density in the following range:
\begin{align}\label{eq:Ofraction}
	f_{\rm decay}\equiv \frac{\Omega_{\rm decay}}{\Omega_{\rm tot}}\simeq 6\%-35\%\, .
\end{align}
It should be emphasized that we have not taken into account the flavor effects from~\eqref{eq:Yukawathermal}. With a single right-handed neutrino mediator, there are in general three coupling constants in~\eqref{eq:Yukawathermal}, corresponding to the   interactions with three lepton flavors. It is possible that all the three couplings are comparably large. In this case, $y_\psi^2= y_{\psi,1}^2+y_{\psi,2}^2+y_{\psi,3}^2$ can further enhance the thermal mass effect within the condition $0.1<y_{\psi,i}<1$. The fraction given in Eq.~\eqref{eq:Ofraction} can then reach a maximal value
$f_{\rm decay}^{\rm max}=43\%$.

A right-handed fermion mediator can also couple to quark doublets. This can be realized by introducing   leptoquarks, which were considered   as   promising candidates to explain flavor anomalies~\cite{Bauer:2015knc,Buttazzo:2017ixm,Li:2022chc}\footnote{Note that, in such leptoquark scenarios, the Yukawa couplings are generically predicted at $\mathcal{O}(1)$.}.  With a scalar leptoquark, the Yukawa interaction is given by
 \begin{align}\label{eq:QLportal}
 	y_{\psi,i}\bar{Q}_{i,L}\psi_{R}	\varphi +\text{h.c.}\, ,
 	%\qquad U_{\mu}\bar{d}\gamma^{\mu}\chi_{R}, 
 \end{align}
where $Q_L=(u_L,d_L)^T$ is the quark doublet and the leptoquark scalar doublet $\varphi$ carries a hypercharge $Y=1/6$.  In this case, the thermal mass of $\psi$ is given by
\begin{align}
	m_\psi^2(T)=\sum_{i=1}^3\frac{3y_{\psi,i}^2}{8}T^2\,,
\end{align}
where the factor of 3 accounts for the color degrees of freedom. If only one coupling is significant, the   condition  $0.1<y_{\psi,i}<1$ would be translated into $0.061<\kappa<0.61$. If  three couplings are comparably large $y_{\psi,1}\approx y_{\psi,2}\approx y_{\psi,3}$, the range of $\kappa$ is given by  $0.11<\kappa<1.1$. In this case, the fraction $f_{\rm decay}$ is estimated to be
\begin{align}
	f_{\rm decay,1f}^{}&=10\%-43\% \, ,
\end{align}
with a  one-flavor (1f) dominated coupling and 
\begin{align}
	f_{\rm decay,3f}^{}&=16\%-45\% \,,
\end{align}
with three-flavor (3f)  comparable couplings.
%-----------------------------------------------------------------------------------------------------------------------
\subsection{Left-handed fermion mediator}
	%-----------------------------------------------------------------------------------------------------------------------
A left-handed   fermion mediator can   couple to the right-handed DM $\chi_R$ via chiral Yukawa interaction. For a nonthermal dark sector via the Yukawa interaction  $\bar \psi_L \chi_R\phi$,  
the left-handed mediator cannot have strong gauge interaction. There are some possibilities. For instance,  $\psi_L$ 
can  couple to the SM charged-lepton singlet $\ell_R$ via~\cite{Bai:2014osa}
\begin{align}\label{eq:eRportal}
y_{\psi,i}\bar{\psi}_{L}\ell_{i,R}\varphi+\text{h.c.}\, ,
\end{align}
where $y_{\psi,i}$ in general have  three  couplings to the  charged-lepton flavors,   $\psi_L$ is a neutral lepton and $\varphi$ is electrically charged. Here $\psi$ is  a SM singlet so that the dark sector does not carry SM gauge charges.  Since both $\varphi$ and $\ell_R$ are SM gauge singlets, the thermal mass of $\psi_L$ would be given by
\begin{align}
	m_\psi^2(T)=\sum_{i=1}^3\frac{ y_{\psi,i}^2}{16}T^2\,,
\end{align}
leading to   
\begin{align}
	f_{\rm decay,1f}^{}=4\%-29\% \,,\quad
	 f_{\rm decay,3f}^{}=7\%-39\% \,,
\end{align}
 for  1f and  3f dominated couplings, respectively.

%------------------------------------------------------------------------------------------------------
\begin{table}
	\centering
	\renewcommand{\arraystretch}{1.4}
	\begin{tabular}{l|c|c}
		\hline\hline
		Thermalization patterns & Range of $f_{\rm decay, 1f}$ (\%) & Range of $f_{\rm decay, 3f}$ (\%) \\ 
		\hline 
		%($\nu+e^-$ scatt.)
		$\bar L_i \varphi \psi_R$& [6,35]           &  [10,43] \\
		\hline	
		$\bar Q_i \varphi \psi_R$& [10,43]           &  [16,45] \\
		\hline
		$\bar\psi_L \varphi \ell_{i,R}$& [4,29]           &  [7,39] \\
		\hline
		$\bar\psi_L \varphi d_{i,R}$& [7,39]           &  [12,45] \\
		\hline	\hline
	\end{tabular}
	\caption{Different thermalization interactions for a right- or left-handed fermion mediator $\psi$ with a Yukawa coupling in the   regime: $[0.1,1]$. 1f assumes that the interaction is dominated by a SM fermion flavor while 3f considers comparable interactions among the three flavors. }
	\label{tab:Ofraction}
\end{table}
%------------------------------------------------------------------------------------------------------

A left-handed fermion mediator singlet can also couple to right-handed quarks. For instance, 
 the   down-quark singlet $d_R$ couples to $\psi_L$ with a leptoquark scalar $\varphi$~\cite{Baker:2015qna,Mandal:2018czf,Baker:2021llj,Belanger:2021smw,Belfatto:2021ats}:
\begin{align}\label{eq:dRportal}
y_{\psi,i}\bar{d}_{i,R}\psi_{L}	\varphi +\text{h.c.}\, ,
	%\qquad U_{\mu}\bar{d}\gamma^{\mu}\chi_{R}, 
\end{align}
where the scalar  $\varphi$ is now an $SU(3)_c$ triplet and $SU(2)_L$ singlet,   carrying the  hypercharge  $Y=-1/3$ so that $\psi$ is a SM singlet.   The thermal mass in this case is given by
\begin{align}
	m_\psi^2(T)=\sum_{i=1}^3\frac{3y_{\psi,i}^2}{16}T^2\,,
\end{align}
where the factor of 3 accounts for the color degrees of freedom. It then leads  to  
\begin{align}
	f_{\rm decay,1f}^{}=7\%-39\% \, ,\quad
	f_{\rm decay,3f}^{}=12\%-45\% \,.
\end{align}

The   thermalization interactions and the portion of forbidden decay are summarized in Tab.~\ref{tab:Ofraction}. A general expectation is that, for a Yukawa coupling at $0.1\textendash 1$, the contribution from the forbidden decay can range from 4\% to   45\%. The largest contribution comes from~\eqref{eq:QLportal} with comparably large Yukawa couplings, where the thermal loop correction to the fermion mediator is enhanced by the gauge degeneracy  and the color degrees of freedom. 

There is no doubt that a phenomenological study of each pattern deserves comprehensive analyses, especially given that they can arise from  crossed  areas, ranging from    neutrino physics, flavor anomalies to   DM physics. The results obtained in this section serve to underlie the detailed investigations when a heavy dark sector is generated by a light fermion mediator.

%-----------------------------------------------------------------------------------------------------------------------
\section{Conclusions}\label{sec:conclusion}
%-----------------------------------------------------------------------------------------------------------------------
In this work we have concentrated on the freeze-in DM production via a light fermion mediator once thermalized in the early universe. We have used   Yukawa portal interactions   to capture the basic properties of such a class of DM models,  where the scattering and forbidden fermion decay rates carry the same order of coupling constants. The results can   be applied   to  the scenarios of right-handed Majorana/Dirac neutrino portals, as well as  the right- and left-handed fermion mediators coupling to  the SM fermions, provided that the dark sector is heavier than the mediator and the relevant thermal particles.

The full forbidden decay rate  should in general be  calculated from the one-loop retarded amplitude at finite temperatures, and is generically overestimated by a tree-level amplitude. Nevertheless, we found that the  forbidden decay rate can still be simply obtained from the tree-level amplitude after being rescaled by  some constants that depend only on the thermal parameter.

Both the scattering and forbidden fermion decay coexist to generate the DM relic density. 
The contribution from the forbidden decay is   significant when the   interaction between the fermion mediator and the thermal plasma is strong.  For a Yukawa coupling in the range: $0.1\textendash 1$, the forbidden decay can contribute to the total DM relic density at the level of $4\%\textendash 45\%$, depending on the gauge representations of thermal particles and flavored Yukawa interactions, and  hence cannot be neglected in    precise calculation  of the  DM relic density.

\begin{acknowledgments}
	\noindent The author thanks  Xun-Jie Xu  for valuable discussions. This work is supported in part by the National Natural Science Foundation of China under grant No. 12141501.
\end{acknowledgments}

\appendix

\section{Thermal one-loop amplitudes}\label{appendix:A}

\subsection{The DM part}\label{appendix:A1}
The amplitudes from Fig.~\ref{fg:SigmaRchi} are given by 
 \begin{align}
 	\Sigma_{+-}^{\chi}(P)	&=-iy_{\chi}^{2}\int\frac{d^{4}K}{(2\pi)^{4}}G_{-+}(K-P)S_{+-}(K)
 	\nonumber \\[0.2cm]
% 	&=-iy_{\chi}^{2}\int\frac{d^{4}K}{(2\pi)^{4}}(-2\pi i)\text{sign}(k_{0}-p_{0})[1+f_{\phi}(k_{0}-p_{0})]\delta[(K-P)^{2}-m_{\phi}^{2}](-2\pi i)f_{\psi}(k_{0})\rho_{\psi}
% 	\nonumber \\[0.2cm]
 	&=\frac{iy_{\chi}^{2}}{(2\pi)^{2}}\int d^{4}K\text{sign}(k_{0}-p_{0})[1+f_{\phi}(k_0-p_0)]f_{\psi}(k_0)\delta_{K-P}\rho_{\psi}(K)\, ,
 	\\[0.3cm]
 		\Sigma_{-+}^{\chi}(P)&=	-iy_{\chi}^{2}\int\frac{d^{4}K}{(2\pi)^{4}}G_{+-}(K-P)S_{-+}(K)
 	\nonumber \\[0.2cm]
 %	&=	-iy_{\chi}^{2}\int\frac{d^{4}K}{(2\pi)^{4}}(-2\pi i)\varepsilon(k_{0}-p_{0})f_{\phi}(k_{0}-p_{0})\delta[(K-P)^{2}-m_{\phi}^{2}](2\pi i)[1-f_{\psi}(k_{0})]\rho_{\psi}
 %	\nonumber \\[0.2cm]
 	&	=	\frac{-iy_{\chi}^{2}}{(2\pi)^{2}}\int d^{4}K\text{sign}(k_{0}-p_{0})f_{\phi}(k_0)[1-f_{\psi}(k_0-p_0)]\delta_{K-P}\rho_{\psi}(K)\, ,
 \end{align}
where $\delta_{K-P}\equiv \delta[(K-P)^{2}-m_{\phi}^{2}]$ and the free scalar propagators $G_{-+}, G_{+-}$ are  given by
\begin{align}\label{eq:G+-}
	G_{+-}(K)&=-2\pi i\text{sign}(k_{0})f_{\phi}(k_{0})\delta(K^{2}-m_{\phi}^{2})\, ,
	%=-2\pi i[\theta(-k_{0})+f_{\phi}(|k_{0}|)]\delta(K^{2}-m_{\phi}^{2})\, ,
	\\[0.3cm]
	G_{-+}(K)&=-2\pi i\text{sign}(k_{0})[1+f_{\phi}(k_{0})]\delta(K^{2}-m_{\phi}^{2})\, ,\label{eq:G-+}
	%=-2\pi i[\theta(k_{0})+f_{\phi}(|k_{0}|)]\delta(K^{2}-m_{\phi}^{2}).
\end{align}
while the resummed fermion propagators $S_{+-}, S_{-+}$ are given by Eqs.~\eqref{eq:S+-} and~\eqref{eq:S-+}.

\subsection{The fermion mediator part}\label{appendix:A2}
The real part of the retarded amplitude $\Sigma_{R}^{\psi}(K)$  is equivalent to the time-ordered one $\Sigma_{++}^{\psi}(K)$, which  in the massless limit is given by
\begin{align}
	\Sigma_{++}^{\psi}(K)	&=iy_{\psi}^{2}\int\frac{d^{4}Q}{(2\pi)^{4}}G_{++}(Q-K)P_{L}S_{++}(Q)P_{R}
	\nonumber \\[0.2cm]
	&=iy_{\psi}^{2}\int\frac{d^{4}Q}{(2\pi)^{4}}\left(\frac{1}{Q^{2}+i\epsilon}+2\pi if_{\eta}(|q_{0}|)\delta(Q^{2})\right)P_{L}\slashed{Q}P_{R}
\nonumber \\[0.2cm]
&	\times\left(\frac{1}{(Q-K)^{2}+i\epsilon}-2\pi if_{\varphi}(|q_{0}-k_{0}|)\delta[(Q-K)^{2}]\right) ,
%	\\[0,5cm]
%\Sigma_{+-}^{\psi}(K)	&=-iy_{\psi}^{2}\int\frac{d^{4}Q}{(2\pi)^{4}}G_{-+}(Q-K)P_{L}S_{+-}(Q)P_{R}
%\nonumber \\[0.2cm]
%&=-iy_{\psi}^{2}\int\frac{d^{4}Q}{(2\pi)^{4}}(-2\pi i)[\theta(q_{0}-k_{0})+f_{S}(|q_{0}-k_{0}|)]\delta[(Q-K)^{2}]
%	\nonumber \\[0.2cm]
%	&\times(2\pi i)[-\theta(-q_{0})+f_{\eta}(|q_{0}|)]\delta(Q^{2})P_{L}\slashed{Q}P_{R}.
\end{align}
The zero-temperature part is UV divergent, which can be renormalized as usual in zero-temperature QFT. For the finite-temperature part, it reads
\begin{align}
 \text{Re}\Sigma_{R}^{\psi}(K)	&=\frac{y_{\psi}^{2}}{(2\pi)^{3}}\int d^{4}Q\Big(\frac{\delta[(Q-K)^{2}]}{Q^{2}}f_{\varphi}(|q_{0}-k_{0}|)-\frac{\delta(Q^{2})}{(Q-K)^{2}}f_{\eta}(|q_{0}|)\Big)P_{L}\slashed{Q}P_{R}
 \nonumber\\[0.2cm]
%	&=\frac{y_{\psi}^{2}}{(2\pi)^{3}}\int d^{4}Q\frac{\delta(Q^{2})}{(Q-K)^{2}}\Big(f_{\varphi}(|q_{0}|)P_{L}(-\slashed{Q}+\slashed{K})P_{R}-f_{\eta}(|q_{0}|)P_{L}\slashed{Q}P_{R}\Big)
% \nonumber\\[0.2cm]
&	=\frac{y_{\psi}^{2}}{(2\pi)^{3}}\int d^4 Q\frac{\delta(Q^{2})}{(Q-K)^{2}}\Big(f_{\varphi}(q)P_{L}(-\slashed{Q}+\slashed{K})P_{R}-f_{\eta}(q)P_{L}\slashed{Q}P_{R}\Big),
\end{align}
where $(Q-K)^{2}\neq0$ and the second equation is obtained by replacing $Q\to-Q+K$ in the first term of the first equation. The above integration  can be done as follows. Integrate  $q_{0}$ first via $\delta(Q^{2})$,  then expand the denominator $(Q-K)^{2}=K^{2}-2K.Q$ in the HTL approximation: $K^{2}\ll q^{2}$\footnote{The forbidden decay   primarily stems from a hard $\psi$ propagating near the lightcone. It implies that  when using the HTL approximation, the terms from  $K^{2}/q^{2}$ have a higher-order   $y_\psi$  but $k_{0}/q$ and $k/q$ are at leading order. },  after that integrate   the angle $\cos\theta$, and finally integrate the momentum $q$.

 In the HTL approximation, the trace given in Eqs.~\eqref{eq:aL} and~\eqref{eq:bL} are evaluated to be  
\begin{align}\label{eq:trKSigma}
	\text{tr}[\slashed{K}\text{Re}\Sigma_{R}^{\psi}(K)]	
	&=\frac{2y_{\psi}^{2}}{(2\pi)^{2}}\int q[f_{\varphi}(q)+f_{\eta}(q)]dq+\mathcal{O}(K^{2}/q^{2})
	\nonumber \\[0.2cm]
	&\approx\frac{y_{\psi}^{2}}{8}T^{2}\, ,
	\\[0.3cm]
	\text{tr}[\slashed{U}\text{Re}\Sigma_{R}^{\psi}(K)]
	&=\frac{y_{\psi}^{2}}{(2\pi)^{2}}\int q[f_{\varphi}(q)+f_{\eta}(q)]dq\int d\cos\theta\frac{k_{0}}{k_{0}^{2}-k^{2}\cos\theta^{2}}+\mathcal{O}(K^{2}/q^{2})
	\nonumber \\[0.2cm]
	&\approx\frac{y_{\psi}^{2}}{16k}\ln\left|\frac{k_{0}+k}{k_{0}-k}\right|T^{2}\, .\label{eq:trUSigma}
\end{align}
%Note that for $\text{tr}[\slashed{U}\text{Re}\Sigma_{R}^{\psi}(K)]$, 
%we have extracted the real part of $\Sigma_{R}^{\psi}(K)$, which indicates that there is no pole in the denominator, so the integrand is the principle value. In fact, in the spacelike region, there is indeed no imaginary part from $\Sigma_{R}^{\psi}(K)$, however, there is imaginary term from $\Sigma_{R}^{\psi}(K)$ in the timelike region. Nevertheless, we move this imaginary term to $\text{Im}\Sigma_{R}^{\psi}(K)$, in which the imaginary term in spacelike region is also included, as will be computed below. To make sure only real term is taken from the integration of angle in $\text{tr}[\slashed{u}\text{Re}\Sigma_{R}^{\psi}(K)]$ for the whole K-space, we add the absolute symbol in the logarithmic function, such that even in the timelike region, only the real part is taken. 

\bibliographystyle{JHEP}
\bibliography{Refs}

\end{document}